\documentclass[aps,preprint,prd,showpacs,nofootinbib,raggedbottom,longbibliography]{revtex4-1}

\pdfoutput=1

\usepackage{amssymb}
\usepackage{appendix}
\usepackage{amsmath}
\usepackage{color}
\usepackage{fancyhdr}
\newcommand{\be}{\begin{equation}}
\newcommand{\ee}{\end{equation}}
\newcommand{\bea}{\begin{eqnarray}}
\newcommand{\eea}{\end{eqnarray}}
\newcommand{\bes}{\begin{subequations}}
\newcommand{\ees}{\end{subequations}}

\newcommand{\bc}{\begin{center}}
\newcommand{\ec}{\end{center}}
\usepackage{amsmath, amssymb} 
\usepackage{slashed}
\usepackage{booktabs}
\usepackage{graphicx}
\usepackage{subcaption}
\usepackage{placeins}
\usepackage{tikz}
\usepackage[compat=1.1.0]{tikz-feynman}
\tikzfeynmanset{
  every fermion={line width=0.8pt},
  every anti fermion={line width=0.8pt},
  every scalar={line width=0.8pt},
  every charged scalar={line width=0.8pt},
  every photon={line width=0.8pt},
  every plain={line width=0.8pt}
}
\usetikzlibrary{positioning}
\usetikzlibrary{shapes.misc} 

\usepackage[colorlinks=true,
            linkcolor=blue,    
            citecolor=blue,    
            urlcolor=blue
           ]{hyperref}

\usepackage{natbib}

\makeatletter
\AtBeginDocument{\global\@namedef{b@apsrev41Control}{{}{}}}
\makeatother

\begin{document}
\title{\(N_{\mathrm{eff}}\), Radiative Dirac Neutrinos, and Flavor Physics from Leptoquarks}

\author{Patricio Escalona$^{1}$}\email{patricioescalona96@gmail.com}
\author{Guillermo Gambini$^{1,2}$}\email{guillerivgamb@gmail.com}
\author{Vitor Pereira$^{1,3}$}\email{vitor.pereira.128@ufrn.edu.br}
\author{C. A. de S. Pires$^{4}$}\email{cpires@fisica.ufpb.br}
\author{Farinaldo S. Queiroz$^{1,5,6}$}\email{farinaldo.queiroz@ufrn.br}

\affiliation{$^1$ International Institute of Physics, Universidade Federal do
Rio Grande do Norte, Campus Universitário, Lagoa Nova, Natal, RN 59078-970,
Brazil}
\affiliation{$^2$ Facultad de Ciencias, Universidad Nacional de Ingeniería
(UNI), Avenida Túpac Amaru 210, Lima, Peru}
\affiliation{$^3$ Departamento de Física, Universidade Federal do Rio Grande
do Norte, Natal, RN 59078-970, Brazil}
\affiliation{$^4$ Departamento de Física, Universidade Federal da Paraíba,
Caixa Postal 5008, João Pessoa, PB 58051-970, Brazil}
\affiliation{$^5$ Millennium Institute for Subatomic Physics at the
High-Energy Frontier (SAPHIR) of ANID, Fernández Concha 700, Santiago, Chile}
\affiliation{$^6$ Departamento de Física, Facultad de Ciencias, Universidad de
La Serena, Avenida Cisternas 1200, La Serena, Chile}
\date{\today}
\begin{abstract}
Leptoquarks provide a common framework for addressing flavor anomalies, rare
meson decays, and neutrino-mass generation. We study a model with two mixed
scalar leptoquarks and three right-handed neutrinos that generates radiative
Dirac neutrino masses and contributes to $B^+\to K^+\nu\bar\nu$,
charged-lepton flavor violation, and the effective number of relativistic
species. We show that the model can accommodate the Belle~II measurement of
$B^+\to K^+\nu\bar\nu$ while satisfying the $\mu\to e\gamma$ bound and the
cosmological constraint on $\Delta N_{\mathrm{eff}}$. When both leptoquark
masses exceed approximately $12$--$14\,\mathrm{TeV}$, the new contribution to
the $B$-meson decay becomes too small, and the prediction approaches the
SM value, which lies below the Belle~II measurement.
Interestingly, the lower bounds on the leptoquark mass scale derived from
$N_{\mathrm{eff}}$ can be stronger than those from direct collider searches.
\end{abstract}

\maketitle
\tableofcontents

\par

\section{Introduction}
Understanding why neutrinos are so much lighter than all other known fermions remains one of the fundamental open questions in particle physics. Although several mechanisms have been proposed to generate tiny neutrino masses, seesaw~\cite{Mohapatra:1979ia,Ma:1998dx} and radiative mechanisms~\cite{Zee:1980ai,Babu:1988ki} are among the most widely explored frameworks.

Another fundamental question in neutrino physics concerns their nature: whether they are Majorana or Dirac particles. The Majorana scenario is particularly appealing from a phenomenological perspective, as it allows for lepton-number-violating processes, with neutrinoless double-beta decay~\cite{Schechter:1981bd} providing its most prominent experimental signature. As a consequence, scenarios in which neutrinos are Majorana particles have been studied more extensively than the Dirac case~\cite{ParticleDataGroup:2026mpi}.

However, the absence of conclusive evidence for neutrinoless double-beta decay, despite decades of experimental searches~\cite{KamLAND-Zen:2024eml,GERDA:2020xhi,CUORE:2021mvw,EXO-200:2019rkq}, has motivated renewed interest in scenarios in which neutrinos are Dirac particles. In this context, seesaw-inspired and radiative mechanisms for generating naturally small Dirac neutrino masses, originally investigated several decades ago~\cite{Chang:1986bp,Mohapatra:1987nx,Babu:1988yq,Lindner:2001hr}, have attracted considerable attention in recent years~\cite{CentellesChulia:2018gwr,CentellesChulia:2018bkz,Oliveira:2025kfg,Calle:2018ovc}.

Whether neutrinos are Dirac or Majorana particles, explaining their tiny masses requires physics beyond the Standard Model (SM). Among well-motivated new-physics scenarios that explain neutrino masses, leptoquark models provide a particularly appealing framework~\cite{Pati:1973uk,Dorsner:2016wpm} in which tiny neutrino masses can arise exclusively through radiative effects~\cite{Mahanta:1999xd,Babu:2010vp,Zhang:2021dgl,Babu:2020hun,Doff:2024ovy, Enkhbat:2013oba,Chen:2022hle}.

Leptoquark couplings to quarks and leptons can induce semileptonic meson decays~\cite{Bauer:2015knc,Becirevic:2016yqi,Bigaran:2019bqv}, including rare $B$-meson decays with neutrinos in the final state, mediated by the flavor-changing transitions $b\to(d,s)\bar\nu\nu$~\cite{Sahoo:2015fla,Calibbi:2017qbu,Chen:2023wpb,Altmannshofer:2023hkn,Crivellin:2025qsq,Chen:2025npb}. Belle~II reported the first evidence for $B^+\to K^+\nu\bar\nu$, with a combined inclusive- and hadronic-tag result at the $10^{-5}$ level, nearly five times the recent SM prediction at the $10^{-6}$ level and corresponding to a $2.7\sigma$ tension~\cite{Belle-II:2023esi,Buras:2024bnu}. This result has motivated several new-physics explanations that can enhance the branching fraction while remaining consistent with other flavor data~\cite{Athron:2023hmz,Fridell:2023ssf,Gabrielli:2024wys,Hou:2024vyw,Chen:2024jlj,Buras:2024ewl}.

Scalar leptoquarks are particularly appealing in this context because their interactions can both modify the underlying $b\to s\nu\bar\nu$ transition and participate in the radiative generation of neutrino masses~\cite{Babu:2020hun,Doff:2024ovy}. Motivated by this connection, we introduce two electroweak-singlet scalar leptoquarks and three right-handed neutrinos, together with a symmetry structure that forbids tree-level Dirac masses and Majorana mass operators. Neutrinos acquire Dirac masses at one loop through down-type quarks and scalar-leptoquark mixing, while the same Yukawa interactions induce flavor-changing neutral currents. We show that this framework can accommodate the $B^+\to K^+\nu\bar\nu$ branching fraction reported by Belle~II while reproducing the measured neutrino mass-squared differences and mixing parameters.

The same leptoquark interactions responsible for neutrino mass generation and rare $B$-meson decays also induce charged-lepton flavor violation, particularly $\mu\to e\gamma$. Moreover, the additional right-handed neutrinos contribute to $\Delta N_{\mathrm{eff}}$ if sufficiently populated in the early Universe. These observables impose complementary constraints on the leptoquark masses and Yukawa couplings. Interestingly, for some Yukawa-coupling configurations, the lower bounds on the leptoquark mass scale from $N_{\mathrm{eff}}$ can be stronger than those from current collider searches.\footnote{This comparison should, however, be regarded as illustrative. Existing ATLAS limits are obtained under simplified assumptions about the leptoquark spectrum, couplings, and decay channels~\cite{ATLAS:2025LQsummary,ATLAS:2025resonant}; a quantitative comparison in the present two-leptoquark framework therefore requires a dedicated collider recast.} We therefore perform a combined phenomenological analysis to identify Yukawa textures and parameter-space regions compatible with neutrino oscillation data and the Belle~II measurement while satisfying constraints from charged-lepton flavor violation and $N_{\mathrm{eff}}$.

This paper is organized as follows. In Sec.~\ref{sec:framework}, we introduce
the model and derive the radiative Dirac neutrino mass matrix.
Section~\ref{sec:perturbativity} discusses the perturbativity requirements on
the Yukawa couplings. Rare meson decays, charged-lepton flavor violation, and
the contribution to $\Delta N_{\mathrm{eff}}$ are studied in
Secs.~\ref{sec:meson-decays}, \ref{sec:clfv}, and \ref{sec:neff}, respectively.
Section~\ref{sec:numerical-analysis} presents the combined numerical analysis
and illustrative benchmarks, and Sec.~\ref{sec:conclusions} summarizes our
conclusions. Appendix~\ref{app:meson-decay-details} provides details of the
meson-decay calculation, while Appendix~\ref{sec:top-decays} discusses rare
top decays.

\section{Framework and neutrino masses}\label{sec:framework}

We extend the SM by introducing three right-handed neutrinos, $\nu_{iR}$, and two independent scalar leptoquarks, $S$ and $\widetilde S$, while keeping the SM charged-fermion content unchanged. We denote the standard left-handed lepton and quark doublets by $L_{iL}$ and $\widehat Q_{iL}$, respectively, with $i=1,2,3$ labeling the fermion generations. Table~\ref{tab:fields} summarizes the fermion and scalar fields, their gauge quantum numbers, and their charges under a discrete $\mathbb Z_2\times\mathbb Z_3$ symmetry imposed to forbid unwanted interactions, particularly tree-level neutrino mass terms and diquark couplings.

\begin{table}[htbp]
\centering
\setlength{\tabcolsep}{12pt}
\renewcommand{\arraystretch}{1.35}
\begin{tabular}{c c c c c c c c}
\hline\hline
Field & $SU(3)_C$ & $SU(2)_L$ & $U(1)_Y$ & $\mathbb Z_2$ & $\mathbb Z_3$ & $B$ & $L$\\ \hline
$\widehat Q_{iL}$ & $\mathbf{3}$ & $\mathbf{2}$ & $\tfrac16$ & $1$ & $1$ & $\tfrac13$ & $0$\\
$\widehat u_{iR}$ & $\mathbf{3}$ & $\mathbf{1}$ & $\tfrac23$ & $1$ & $1$ & $\tfrac13$ & $0$\\
$\widehat d_{iR}$ & $\mathbf{3}$ & $\mathbf{1}$ & $-\tfrac13$ & $1$ & $1$ & $\tfrac13$ & $0$\\
$L_{iL}$ & $\mathbf{1}$ & $\mathbf{2}$ & $-\tfrac12$ & $-1$ & $1$ & $0$ & $1$\\
$e_{iR}$ & $\mathbf{1}$ & $\mathbf{1}$ & $-1$ & $-1$ & $1$ & $0$ & $1$\\
$\nu_{iR}$ & $\mathbf{1}$ & $\mathbf{1}$ & $0$ & $1$ & $\omega_3^2$ & $0$ & $1$\\
$H$ & $\mathbf{1}$ & $\mathbf{2}$ & $\tfrac12$ & $1$ & $1$ & $0$ & $0$\\
$S$ & $\overline{\mathbf{3}}$ & $\mathbf{1}$ & $\tfrac13$ & $-1$ & $1$ & $-\tfrac13$ & $-1$\\
$\widetilde S$ & $\overline{\mathbf{3}}$ & $\mathbf{1}$ & $\tfrac13$ & $1$ & $\omega_3$ & $-\tfrac13$ & $-1$\\ \hline\hline
\end{tabular}
\caption{Field content and gauge, discrete, baryon, and lepton charges.
We use the hypercharge convention $Q=T_3+Y$ and define $\omega_3=e^{2\pi i/3}$. The generation index is $i=1,2,3$. Hats denote quark weak eigenstates; $S$ and $\widetilde S$ are
independent scalar leptoquarks.}
\label{tab:fields}
\end{table}

The SM Higgs doublet is parametrized as $H=(G^+,(v+h+iG^0)/\sqrt{2})^T$, where $v$ denotes the electroweak vacuum expectation value, $h$ is the physical Higgs boson, and $G^\pm$ and $G^0$ are the would-be Goldstone bosons associated with electroweak symmetry breaking.

The Yukawa interactions consistent with the particle content and all symmetries of the model are given by
\begin{align}
\mathcal L^Y={}&
 y^l_{ii}\overline L_{iL}e_{iR}H
 +y^d_{ij}\overline{\widehat Q}_{iL}\widehat d_{jR}H
 +y^u_{ij}\overline{\widehat Q}_{iL}\widetilde H\widehat u_{jR}
 \nonumber\\
&+Y^L_{i\alpha}\overline{\widehat Q^{Ca}_{iL}}
 \epsilon^{ab}L^b_{\alpha L}S
 +Y^R_{i\alpha}\overline{\widehat d^C_{iR}}\nu_{\alpha R}\widetilde S
 +\widetilde Y^R_{i\alpha}\overline{\widehat u^C_{iR}}e_{\alpha R}S
 +\mathrm{H.c.},
\label{eq:alt-yukawa}
\end{align}
where $\widetilde H=i\sigma_2H^*$, $\epsilon^{12}=1$, hats denote
quark weak eigenstates, and $\alpha$ labels lepton flavors.

This set of interactions conserves baryon and lepton
numbers with $B(S)=B(\widetilde S)=-1/3$ and
$L(S)=L(\widetilde S)=-1$. This conservation persists after the soft
scalar mixing introduced below, protecting proton stability and the
Dirac nature of neutrinos within the renormalizable theory. 

We choose the charged-lepton mass basis, so $y^l$ is diagonal, and a
quark weak basis with trivial right-handed quark rotations. The weak and
mass eigenstates of the left-handed up- and down-type quarks are related by
\begin{equation}
\widehat u_L=V_L^u u_L,\qquad
\widehat d_L=V_L^d d_L.
\label{eq:alt-quark-basis}
\end{equation}
The matrices $V_L^u$ and $V_L^d$ are unitary, and their mismatch defines the
Cabibbo--Kobayashi--Maskawa (CKM) matrix through
$V\equiv V_{\mathrm{CKM}}=(V_L^u)^\dagger V_L^d$. In the standard CKM
parametrization, we use $s_{12}=0.22517$, $s_{13}=0.003763$,
$s_{23}=0.04189$, and $\delta=1.154$, as reported by the 2026
PDG~\cite{ParticleDataGroup:2026mpi}. We work in the convenient basis
$V_L^u=I$, so that $V_L^d=V$, and independently assume that $Y^R$ is
diagonal. Since
$\overline{\widehat q^{\,C}}=\overline{q^C}(V_L^q)^T$, the
leptoquark interactions in the fermion mass basis, with the left-handed
neutrinos still labeled by weak flavor, are
\begin{equation}
\mathcal L_{\mathrm{LQ}}^Y\supset Y^L_{i\alpha}\overline{u^C_{iL}}e_{\alpha L}S-\zeta_{i\alpha}\overline{d^C_{iL}}\nu_{\alpha L}S+Y^R_{ii}\overline{d^C_{iR}}\nu_{iR}\widetilde S+\widetilde Y^R_{i\alpha}\overline{u^C_{iR}}e_{\alpha R}S+\mathrm{H.c.}
\label{eq:alt-physical-yukawa}
\end{equation}
The rotation of the down-type quarks defines $\zeta\equiv V^T Y^L$, with
the inverse relation $Y^L=V^*\zeta$. Here $u_{1,2,3}=u,c,t$ and
$d_{1,2,3}=d,s,b$.
Although allowed by the symmetries, the coupling $\widetilde Y^R$ does not
enter the one-loop mechanism responsible for the Dirac neutrino masses.

The scalar potential consistent with the symmetries of the model is
\begin{align}
V_0={}&\mu^2H^\dagger H+\mu_S^2S^\dagger S
 +\mu_{\widetilde S}^2\widetilde S^\dagger\widetilde S
 +\lambda_1(H^\dagger H)^2
 +\lambda_2(S^\dagger S)^2
 +\lambda_3(\widetilde S^\dagger\widetilde S)^2
 \nonumber\\
&+\lambda_4(H^\dagger H)(S^\dagger S)
 +\lambda_5(H^\dagger H)(\widetilde S^\dagger\widetilde S)
 +\lambda_6(S^\dagger S)(\widetilde S^\dagger\widetilde S)
 +\lambda_7|S^\dagger\widetilde S|^2.
\label{eq:alt-potential}
\end{align}
To preserve color, we adopt the standard electroweak vacuum,
$\langle H\rangle=(0,v/\sqrt{2})^T$ and
$\langle S\rangle=\langle\widetilde S\rangle=0$.
The Higgs mass is $m_h^2=2\lambda_1v^2$, and the leptoquark diagonal
mass parameters are
$m_S^2=\mu_S^2+\lambda_4v^2/2$ and
$m_{\widetilde S}^2=\mu_{\widetilde S}^2+\lambda_5v^2/2$.
To generate neutrino masses, we introduce a soft symmetry-breaking term that
mixes the two leptoquarks and induces a non-diagonal mass matrix,
\begin{equation}
V_{\mathrm{soft}}=\mu_{S\widetilde S}^2S^\dagger\widetilde S
 +\mathrm{H.c.},\qquad
\mathcal M_S^2=
\begin{pmatrix}
m_S^2&\mu_{S\widetilde S}^2\\
\mu_{S\widetilde S}^2&m_{\widetilde S}^2
\end{pmatrix}.
\label{eq:alt-scalar-mass}
\end{equation}
The mass matrix $\mathcal M_S^2$ is diagonalized by an orthogonal rotation
with scalar mixing angle $\theta$,
\begin{equation}
\begin{pmatrix}S_1\\S_2\end{pmatrix}
=\begin{pmatrix}c_\theta&s_\theta\\-s_\theta&c_\theta\end{pmatrix}
\begin{pmatrix}S\\\widetilde S\end{pmatrix},\qquad
\sin(2\theta)=\frac{2\mu_{S\widetilde S}^2}
 {m_{S_1}^2-m_{S_2}^2},
\label{eq:alt-scalar-rotation}
\end{equation}
where $c_\theta=\cos\theta$ and $s_\theta=\sin\theta$.
All interactions with physical leptoquarks follow from
Eq.~\eqref{eq:alt-physical-yukawa} by replacing
$S=c_\theta S_1-s_\theta S_2$ and
$\widetilde S=s_\theta S_1+c_\theta S_2$.

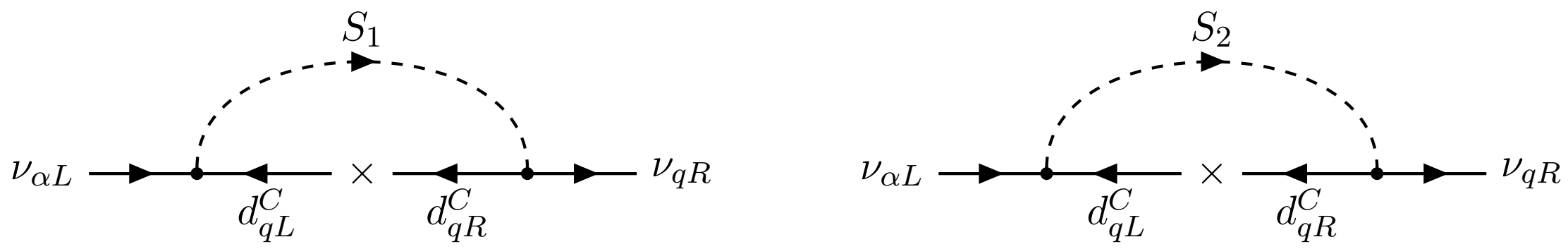
\begin{figure}[t!]
\centering
\setlength{\tabcolsep}{16pt}
\begin{tabular}{@{}cc@{}}
\begin{tikzpicture}[
 baseline=(current bounding box.center),
 every node/.style={font=\small}
]
\begin{feynman}
 \vertex (nL) at (-3.0,0) {$\nu_{\alpha L}$};
 \vertex[dot, minimum size=3pt] (vL) at (-1.55,0) {};
 \vertex (mx) at (0,0) {$\times$};
 \vertex[dot, minimum size=3pt] (vR) at (1.55,0) {};
 \vertex (nR) at (3.0,0) {$\nu_{qR}$};
 \diagram*{
  (nL) -- [fermion] (vL),
  (vL) -- [anti fermion, edge label'={$d^C_{qL}$}] (mx),
  (mx) -- [anti fermion, edge label'={$d^C_{qR}$}] (vR),
  (vR) -- [fermion] (nR),
  (vL) -- [charged scalar, half left, looseness=1.10,
           edge label={$S_1$}] (vR)
 };
\end{feynman}
\end{tikzpicture}
&
\begin{tikzpicture}[
 baseline=(current bounding box.center),
 every node/.style={font=\small}
]
\begin{feynman}
 \vertex (nL) at (-3.0,0) {$\nu_{\alpha L}$};
 \vertex[dot, minimum size=3pt] (vL) at (-1.55,0) {};
 \vertex (mx) at (0,0) {$\times$};
 \vertex[dot, minimum size=3pt] (vR) at (1.55,0) {};
 \vertex (nR) at (3.0,0) {$\nu_{qR}$};
 \diagram*{
  (nL) -- [fermion] (vL),
  (vL) -- [anti fermion, edge label'={$d^C_{qL}$}] (mx),
  (mx) -- [anti fermion, edge label'={$d^C_{qR}$}] (vR),
  (vR) -- [fermion] (nR),
  (vL) -- [charged scalar, half left, looseness=1.10,
           edge label={$S_2$}] (vR)
 };
\end{feynman}
\end{tikzpicture}
\end{tabular}
\caption{One-loop Dirac-neutrino mass diagrams for diagonal $Y^R$.
The external indices are the left-handed weak flavor $\alpha$ and the
right-handed generation $q$, which also labels the internal down quark.
The cross denotes its mass insertion; the two scalar contributions
enter with opposite mixing factors.}
\label{fig:one-loop-observables}
\end{figure}

The Yukawa interactions in Eq.~\eqref{eq:alt-physical-yukawa}, together with mixing between the scalar leptoquarks $S$ and $\widetilde S$, generate Dirac neutrino masses at one loop, as illustrated in Fig.~\ref{fig:one-loop-observables}. The required chirality flip occurs along the internal down-type quark line, rendering the resulting neutrino masses proportional to the corresponding quark masses. For diagonal
$Y^R$, the result is~\cite{Farzan:2012sa}
\begin{align}
 (m_\nu^D)_{\alpha q}= \frac{3\sin(2\theta)}{32\pi^2}\, \zeta_{q\alpha}^*m_{d_q}Y^R_{qq}
 \left[F(m_{S_1}^2,m_{d_q}^2)-F(m_{S_2}^2,m_{d_q}^2)\right],
\label{eq:alt-neutrino-mass}
\end{align}
with no sum over $q$ and $(m_{d_1},m_{d_2},m_{d_3})=(m_d,m_s,m_b)$. The factor of three accounts for the three quark colors. The loop function is
\begin{equation}
F(M^2,m^2)=\frac{M^2}{M^2-m^2}\ln\frac{M^2}{m^2}.
\label{eq:alt-loop-function}
\end{equation}
In the heavy-leptoquark limit, $m_{S_1},m_{S_2}\gg m_{d_q}$, the difference
of loop functions becomes
\begin{equation}
F(m_{S_1}^2,m_{d_q}^2)-F(m_{S_2}^2,m_{d_q}^2)
 \simeq\ln\frac{m_{S_1}^2}{m_{S_2}^2}.
\label{eq:alt-loop-function-heavy-limit}
\end{equation}
Thus, at finite Yukawa couplings, neutrino masses vanish for zero mixing or
exact scalar degeneracy.

In matrix form, we have
\begin{align}
m_\nu^D={}&\zeta^\dagger\mathcal A,\,\,\,\,\mbox{with}\,\,\,
\mathcal A=\frac{3\sin(2\theta)}{32\pi^2}\,m_{d}Y^R
 \left[F(m_{S_1}^2,m_{d}^2)-F(m_{S_2}^2,m_{d}^2)\right].
\label{eq:alt-neutrino-mass-matrix}
\end{align}
Here $m_d$ is the diagonal down-quark mass matrix, the loop function $F$ is evaluated separately for each diagonal mass-squared entry $m_{d_q}^2$, and $\mathcal A_q\equiv(\mathcal A)_{qq}$.
The Dirac mass matrix admits the biunitary decomposition~\cite{Ludl:2014axa}
\begin{equation}
m_\nu^D=U_{\mathrm{PMNS}}D_\nu V_R^\dagger,\qquad
D_\nu=\operatorname{diag}(m_1,m_2,m_3).
\label{eq:alt-diagonalization}
\end{equation}
When $\mathcal A$ is nonsingular, all left-handed leptoquark couplings
follow from
\begin{equation}
\zeta=(\mathcal A^{-1})^\dagger V_R D_\nu U_{\mathrm{PMNS}}^\dagger \qquad \to
\qquad
Y^L=V^*(\mathcal A^{-1})^\dagger V_R D_\nu U_{\mathrm{PMNS}}^\dagger.
\label{eq:alt-reconstruction}
\end{equation}

For the oscillation inputs, we use the NuFIT~6.0 best fit including
atmospheric information~\cite{Esteban:2024nufit}:
\begin{equation}
\begin{alignedat}{2}
\sin^2\theta_{12} &= 0.308,
&\qquad \delta_{\mathrm{CP}} &= 212^\circ, \\
\sin^2\theta_{23} &= 0.470,
&\qquad \Delta m_{21}^2 &= 7.49\times10^{-5}\,\mathrm{eV}^2, \\
\sin^2\theta_{13} &= 0.02215,
&\qquad \Delta m_{31}^2 &= 2.513\times10^{-3}\,\mathrm{eV}^2.
\end{alignedat}
\label{eq:alt-oscillation-inputs}
\end{equation}
The quoted values correspond to normal ordering~\cite{Esteban:2024nufit}, which we adopt in
the numerical analysis. We fix $m_1=0$, so
$m_2=\sqrt{\Delta m_{21}^2}$ and $m_3=\sqrt{\Delta m_{31}^2}$,
and retain the PMNS phase. We additionally choose $V_R=I$, aligning
the right-handed neutrino mass basis with the basis in which $Y^R$
is diagonal. The first column of $m_\nu^D$ must then vanish:
\begin{equation}
(m_\nu^D)_{\alpha1}=\zeta_{1\alpha}^*\mathcal A_1=0
\qquad\text{for every }\alpha.
\label{eq:alt-massless-limit}
\end{equation}
For nondegenerate scalars and nonzero $\sin(2\theta)$, this requires
either $Y^R_{11}=0$ or $\zeta_{1\alpha}=0$ for all $\alpha$, but does
not require both. If $Y^R_{11}\ne0$, the first row of $\zeta$ must
vanish; if $Y^R_{11}=0$, that row is unconstrained by the mass relation.

For the combined numerical analysis in Sec.~\ref{sec:numerical-analysis}, we
adopt the simple texture
\begin{equation}
Y^R=\operatorname{diag}(0,y_2,y_3),\qquad y_2,y_3>0,\qquad
\zeta_{1\alpha}=0,\qquad \widetilde Y^R=0.
\label{eq:alt-numerical-texture}
\end{equation}
Thus we set both factors in Eq.~\eqref{eq:alt-massless-limit} to zero
as an additional simplifying choice. Since $\mathcal A_1=0$, the inverse in
Eq.~\eqref{eq:alt-reconstruction} is used only on $q=2,3$, giving
\begin{equation}
\zeta_{q\alpha}=\frac{m_{\nu_q} U_{\alpha q}^*}{\mathcal A_q^*}\quad(q=2,3),
\qquad Y^L=V^*\zeta.
\label{eq:alt-singular-texture}
\end{equation}
Here, for $q=2,3$, $m_{\nu_q}$ denotes the $q$th neutrino mass, whereas
$m_{d_q}$ denotes the corresponding down-quark mass. The CKM rotation generally
leaves the first row of $Y^L$ nonzero even though $\zeta_{1\alpha}=0$.

Thus, the phenomenology is determined by the scalar masses, the scalar mixing angle,
and the Yukawa couplings in Eq.~\eqref{eq:alt-physical-yukawa}. Using a
diagonal $Y^R$ as input, the reconstruction in
Eqs.~\eqref{eq:alt-reconstruction} and~\eqref{eq:alt-singular-texture}
expresses $\zeta$ and $Y^L$ in terms of $Y^R$ and the neutrino oscillation and
quark mixing inputs. In the following section, we study the perturbativity of
the Yukawa couplings within this texture.

\section{Perturbativity of Yukawa couplings}
\label{sec:perturbativity}
We apply the texture in Eq.~\eqref{eq:alt-numerical-texture} and the
reconstruction in Eq.~\eqref{eq:alt-singular-texture}. We apply the conventional elementwise perturbativity cutoff
$\sqrt{4\pi}$ to the Yukawa matrices $Y^R$, $\zeta$, and $Y^L$,
following Ref.~\cite{Khasianevich:2023duu}.
For the chosen texture, this condition reads
\begin{equation}
\Lambda(y_2,y_3)\equiv
\max\left\{y_2,y_3,
 \max_{\substack{q=2,3\\\alpha}}|\zeta_{q\alpha}|,
 \max_{i,\alpha}|Y^L_{i\alpha}|\right\}
<\sqrt{4\pi}.
\label{eq:alt-perturbativity-condition}
\end{equation}
We use Eq.~\eqref{eq:alt-perturbativity-condition} as a selection
criterion for the numerical scan. Partial-wave unitarity bounds
on Yukawa couplings depend on the scattering channels and field
representations~\cite{Allwicher:2021rtd}. Through the reconstruction in Eq.~\eqref{eq:alt-singular-texture}, each entry
of $Y^L$ generally depends on both $y_2$ and $y_3$. The full criterion
therefore requires a joint test of the two couplings.

In this section, we evaluate $\mathcal A_q$ in
Eq.~\eqref{eq:alt-neutrino-mass-matrix} using the logarithmic approximation
of Eq.~\eqref{eq:alt-loop-function-heavy-limit}, valid for
$m_{S_1},m_{S_2}\gg m_b$. The scalar dependence is then contained in
$\sin(2\theta)$ and the mass ratio. It is useful to define the relative
scalar-mass splitting
$\Delta_S\equiv2|m_{S_1}-m_{S_2}|/(m_{S_1}+m_{S_2})$, with
$0\leq\Delta_S<2$. The logarithmic factor can then be written as
$|\ln(m_{S_1}^2/m_{S_2}^2)|=4\operatorname{arctanh}(\Delta_S/2)$.

\begin{figure}[t!]
\centering
\includegraphics[width=0.78\textwidth]{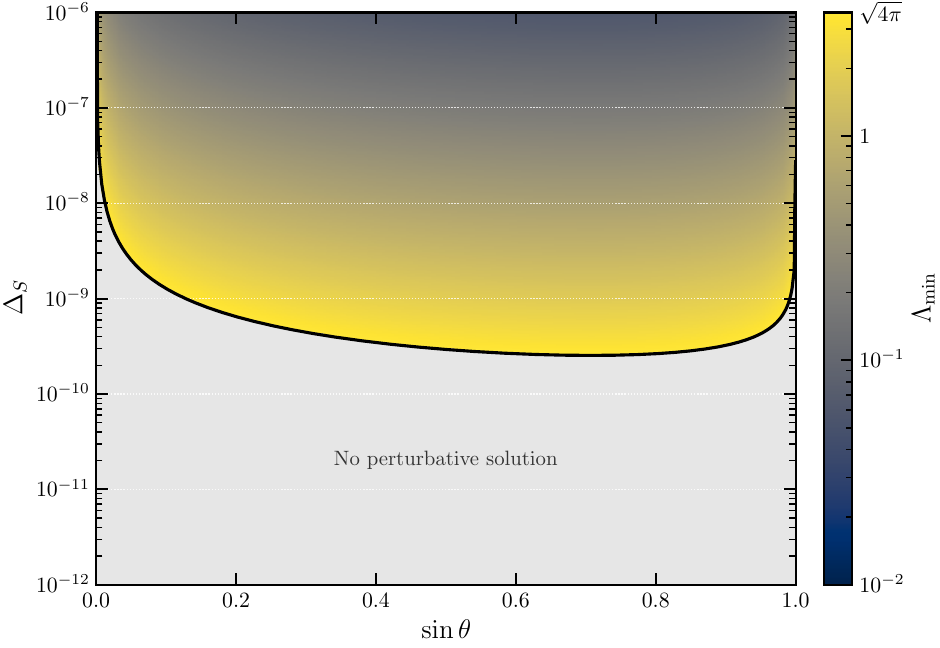}
\caption{Perturbativity map for the texture in
Eq.~\eqref{eq:alt-numerical-texture}. The color scale gives $\Lambda_{\min}$
where a perturbative texture exists; light gray denotes no perturbative
solution for the common bound on $Y^R$, $\zeta$, and $Y^L$. The black contour marks
$\Lambda_{\min}=\sqrt{4\pi}$.}
\label{fig:alt-perturbativity-log-splitting}
\end{figure}

Note that, at fixed $(\sin\theta,\Delta_S)$, increasing an entry $y_q$ of
$Y^R$ suppresses the corresponding row of $\zeta$ and its contribution to
$Y^L$, while decreasing $y_q$ has the opposite effect. These competing
tendencies constrain the allowed values of $y_2$ and $y_3$. At each point in the
$(\sin\theta,\Delta_S)$ plane, we ask whether at least one pair satisfies
the common perturbativity bound. This existence condition can be written as
\begin{equation}
\Lambda_{\min}(\sin\theta,\Delta_S)
=\min_{y_2,y_3>0}\Lambda(y_2,y_3),\qquad
\Lambda_{\min}<\sqrt{4\pi}.
\label{eq:alt-perturbativity-minimum}
\end{equation}
For nondegenerate scalar masses and nonzero $\sin(2\theta)$, the
reconstruction in Eq.~\eqref{eq:alt-singular-texture} reproduces
the input neutrino masses and mixing for each positive pair
$(y_2,y_3)$. Equation~\eqref{eq:alt-perturbativity-minimum}
tests whether any of these pairs yields perturbative couplings. Fig.~\ref{fig:alt-perturbativity-log-splitting}
shows $\Lambda_{\min}$ on the color scale in the perturbative region.
Light gray indicates points with no perturbative solution, and the black
contour marks the boundary $\Lambda_{\min}=\sqrt{4\pi}$.

At exact scalar degeneracy or either mixing endpoint,
$\sin\theta=0$ or $1$, the radiative mass vanishes for finite Yukawa
couplings, so no texture reproduces $m_2,m_3\ne0$. Approaching these
limits suppresses the loop factor and requires larger Yukawa couplings
to maintain the neutrino masses. The joint criterion above quantifies
the corresponding loss of perturbative solutions.

\section{Rare meson decays}\label{sec:meson-decays}
\begin{figure}[t!]
\centering
\begin{tikzpicture}[every node/.style={font=\small}]
  \begin{feynman}
    \vertex (b) at (-2.8,0) {$\bar b_A$};
    \vertex[dot, minimum size=3pt] (vL) at (-1.3,0) {};
    \vertex[dot, minimum size=3pt] (vR) at (0.6,0) {};
    \vertex (nubar) at (2.6,1.35) {$\bar\nu_{jA}$};
    \vertex (nu) at (2.6,0.70) {$\nu_{iB}$};
    \vertex (s) at (2.6,0) {$\bar s_B$};
    \vertex (uspin) at (-2.8,-0.65) {$u$};
    \vertex (uspout) at (2.6,-0.65) {$u$};
    \diagram*{
      (b) -- [anti fermion] (vL),
      (vL) -- [anti fermion] (nubar),
      (vL) -- [scalar] (vR),
      (vR) -- [anti fermion] (s),
      (vR) -- [fermion] (nu),
      (uspin) -- [fermion] (uspout)
    };
  \end{feynman}
  \node[fill=white,inner sep=1.5pt] at (-0.35,-0.32) {$S_a$};
  \node at (-3.55,-0.32) {$B^+$};
  \node at (3.35,-0.32) {$K^+$};
\end{tikzpicture}
\caption{Tree-level contribution to $B^+\to K^+\nu_i\bar\nu_j$ from
the physical leptoquarks $S_a$ ($a=1,2$). The choices $A,B=L,R$
include both equal-chirality vector and mixed-chirality scalar and tensor
contributions.}
\label{fig:meson-lq-exchange}
\end{figure}
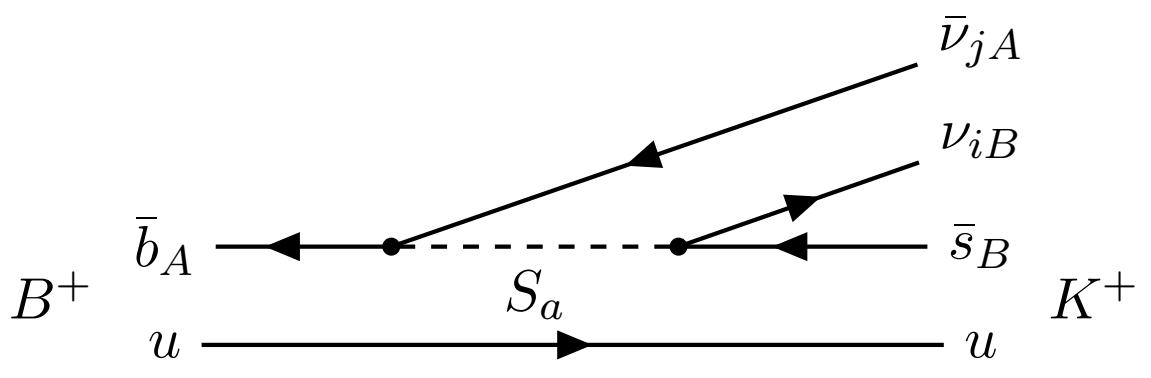

At the quark level, $B^+\to K^+\nu\bar\nu$ proceeds through
$\bar b\to\bar s\nu\bar\nu$, the charge conjugate of
$b\to s\nu\bar\nu$. 
Combining its inclusive- and hadronic-tag analyses, Belle~II obtained~\cite{Belle-II:2023esi}
\begin{equation}
    \mathcal{B}(B^+ \to K^+\nu\bar\nu)
    =\left[2.3\pm0.5\,(\mathrm{stat})^{+0.5}_{-0.4}\,(\mathrm{syst})\right]
    \times10^{-5}.
\label{eq:belle-branching-ratio}
\end{equation}
Combining the statistical and systematic uncertainties in quadrature
separately on each side gives the experimental $1\sigma$ interval
$\mathcal{B}(B^+\to K^+\nu\bar\nu)\in
[1.6596876,3.0071068]\times10^{-5}$.
A recent short-distance SM prediction based on the HPQCD form factors
and an inclusive determination of $|V_{cb}|$ gives~\cite{Buras:2024bnu}
\begin{equation}
 \mathcal{B}(B^+ \to K^+\nu \bar \nu)_{\mathrm{SM}}^{\mathrm{SD}} =
(4.78 \pm 0.30) \times 10^{-6}.
\label{eq:sm-branching-ratio}
\end{equation}
Our numerical implementation is consistent with this result.\footnote{Without
leptoquark contributions, we obtain
$\mathcal B_{\mathrm{SM}}=4.79873\times10^{-6}$, only $0.4\%$ above the quoted
central value and well within its uncertainty; the difference reflects the
specific CKM and electroweak inputs and the numerical form-factor integration.}
Thus, the Belle~II central value is approximately $4.8$ times the SM
prediction.
Fig.~\ref{fig:meson-lq-exchange} shows the tree-level leptoquark
contribution: the $\bar b$ transitions to a $\bar s$, while the $u$
quark is a spectator.

\newcommand{\lqtreediagram}[4]{%
\begin{tikzpicture}[baseline=(current bounding box.center),
    every node/.style={font=\small}]
  \begin{feynman}
    \vertex (in) at (-3.0,0) {$#1$};
    \vertex (outL) at (2.6,1.2) {$#2$};
    \vertex[dot, minimum size=3pt] (vL) at (-1.3,0) {};
    \vertex[dot, minimum size=3pt] (vR) at (0.6,-0.8) {};
    \vertex (outR) at (2.6,-0.2) {$#3$};
    \vertex (outB) at (2.6,-1.6) {$#4$};
    \diagram*{
      (in) -- [fermion] (vL),
      (vL) -- [anti fermion] (outL),
      (vL) -- [scalar, edge label={$S_a$}] (vR),
      (vR) -- [fermion] (outR),
      (vR) -- [fermion] (outB)
    };
  \end{feynman}
\end{tikzpicture}%
}

Since the neutrino flavors are not resolved, we sum the differential
rate, including the SM and leptoquark contributions, over all
$\nu_i\bar\nu_j$ final states, including $i\ne j$.
Integrating this inclusive rate over the allowed range of the
neutrino-pair invariant mass squared gives
\begin{equation}
 \mathcal{B}(B^+\to K^+\nu\bar\nu)=
 \tau_{B^+}\int_0^{(m_B-m_K)^2}
 \frac{d\Gamma}{dq^2}\, dq^2,
\label{eq:BR}
\end{equation}
where $m_B$ and $m_K$ are the charged-meson masses,
$q^2$ is the invariant mass squared of the neutrino pair, and
$\tau_{B^+}=(1.638\pm0.004)\times10^{-12}\,\mathrm{s}$ is the charged-$B$
lifetime~\cite{ParticleDataGroup:2026mpi}.

The down-quark-neutrino couplings of the physical leptoquarks are
\begin{equation}
\mathcal{Y}_L^{(1)}=c_\theta\,\zeta,\hspace{0.1in} \mathcal{Y}_L^{(2)}=-s_\theta\, \zeta,\hspace{0.1in} \mathcal{Y}_R^{(1)}=s_\theta\, Y^R,\hspace{0.1in} \mathcal{Y}_R^{(2)}=c_\theta\, Y^R.
\label{eq:meson-physical-yukawa}
\end{equation}
Here $\zeta$ and $Y^R$ are defined in
Eq.~\eqref{eq:alt-physical-yukawa}. At tree level, two vertices of the
same chirality generate vector operators, while opposite chiralities
generate scalar and tensor operators. We evolve the contribution of each
scalar from its respective matching scale down to $\mu_b=4.8\,\mathrm{GeV}$
and employ the HPQCD lattice-QCD form factors~\cite{Parrott:2022rgu,Parrott:2022sm}. The differential rate,
matching coefficients, QCD evolution, and form-factor inputs are given
in Appendix~\ref{app:meson-decay-details}.

Fig.~\ref{fig:three_plots} shows the branching fraction for
$\sin\theta=10^{-9}$, $3\times10^{-9}$, and $0.1$, with
$Y^R=\operatorname{diag}(1,1,1)$. For these choices, points compatible with
Belle~II occur mainly for $m_{S_2}$ of a few TeV, while $m_{S_1}$ can extend
into the tens-of-TeV range. As scalar degeneracy is approached, the
loop-function difference in Eq.~\eqref{eq:alt-neutrino-mass} vanishes, and
reproducing the neutrino masses requires increasingly large Yukawa couplings.
The gray region identifies points that fail the perturbativity criterion in
Eq.~\eqref{eq:alt-perturbativity-condition}. This region narrows as the mixing
angle increases across the three panels because, at fixed $Y^R$ and scalar
masses, the required left-handed couplings scale as
$1/|\sin(2\theta)|$.

\begin{figure}[t!]
    \centering
    \includegraphics[width=\linewidth]{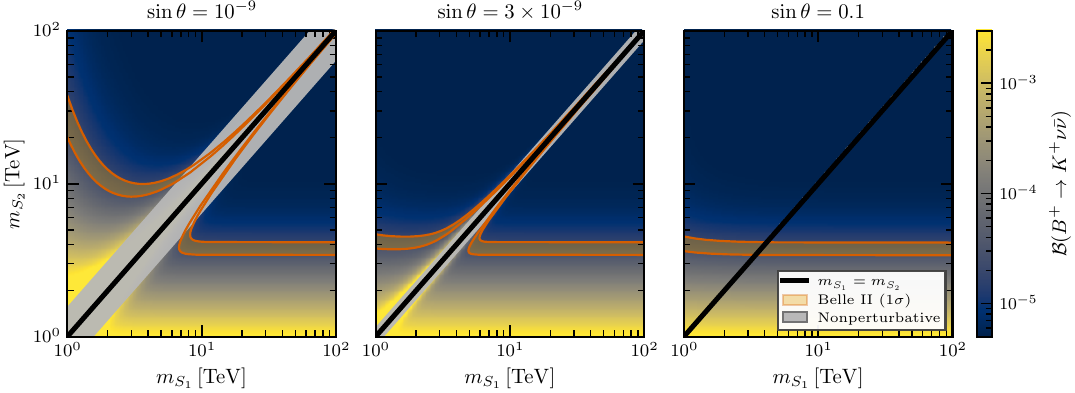}
    \caption{Branching fraction for $B^+\to K^+\nu\bar\nu$ in the
    $(m_{S_1},m_{S_2})$ plane for $Y^R=\operatorname{diag}(1,1,1)$.
    The panels show $\sin\theta=10^{-9}$ (left), $3\times10^{-9}$ (center),
    and $0.1$ (right). The black line indicates $m_{S_1}=m_{S_2}$.
    In the gray region, at least one entry of $Y^R$, $\zeta$, or
    $Y^L=V^*\zeta$ reaches or exceeds $\sqrt{4\pi}$. The orange band is
    the experimental $1\sigma$ interval from the Belle~II measurement in
    Eq.~\eqref{eq:belle-branching-ratio}, with statistical and systematic
    uncertainties combined in quadrature.}
    \label{fig:three_plots}
\end{figure}

\section{Charged-lepton flavor violation}\label{sec:clfv}

\begin{figure}[t!]
    \centering
    \includegraphics[width=0.96\linewidth]{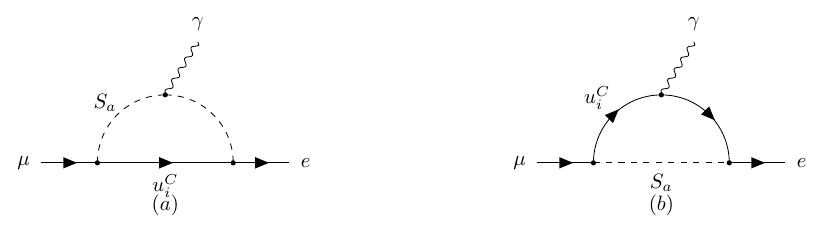}
    \caption{One-loop diagrams contributing to the electromagnetic dipole
    transition $\mu\to e\gamma$. The photon is emitted from the physical
    leptoquark $S_a$ in the left panel and from the charge-conjugate up-type
    quark $u_i^C$ in the right panel. Here $a=1,2$.}
    \label{fig:electromagnetic-dipole-transitions}
\end{figure}

Charged-lepton flavor violation (CLFV) provides a particularly clean probe of
the leptoquark Yukawa sector. In the SM extended only to include light neutrino
masses and mixing, the predicted branching fraction for
$\mu\to e\gamma$ is of order $10^{-54}$~\cite{Lee:1977tib,Palo:2023irb,Calibbi:2017uvl},
far below experimental sensitivity. Consequently, any
observable signal would constitute unambiguous evidence for physics beyond
the SM. In the present model, the same couplings $Y^L$
reconstructed in Eq.~\eqref{eq:alt-reconstruction} couple charged leptons to
up-type quarks and generate the electromagnetic dipole transition at one loop,
thereby testing the flavor structure required to reproduce neutrino
oscillation data.
Fig.~\ref{fig:electromagnetic-dipole-transitions} shows the relevant diagrams.

For the decay $\mu \to e\gamma$, the branching fraction is~\cite{Lavoura:2003xp}\footnote{We use the heavy-leptoquark limit,
$m_{u_i}^2/m_{S_a}^2\ll1$, neglecting quark masses in the loop functions.
The $100$--$150\,\mathrm{GeV}$ results in
Fig.~\ref{fig:numerical-cumulative} are therefore approximate and can be
refined by retaining the full top-quark-mass dependence.}
\begin{equation}\label{eq:br-muegamma}
\mathcal{B}(\mu\to e\gamma)
\simeq
\frac{3\alpha_{\mathrm{em}}}
{256\pi G_F^2}
\left|
\left(
\frac{\cos^2\theta}{m_{S_1}^2}
+
\frac{\sin^2\theta}{m_{S_2}^2}
\right)
\sum_{i=1}^{3}
Y^L_{i2}Y^{L*}_{i1}
\right|^2.
\end{equation}

Here $\alpha_{\mathrm{em}}$ is the electromagnetic fine-structure constant and
$G_F$ is the Fermi constant. The flavor-violating combination
$\sum_i Y^L_{i2}Y^{L*}_{i1}$ is fixed once the neutrino masses and mixings, the
right-handed Yukawa matrix $Y^R$, and the scalar-sector parameters are chosen.
Since $Y^{L\dagger}Y^L=\zeta^\dagger\zeta$, this flavor factor is independent
of the CKM rotation in the approximation of Eq.~\eqref{eq:br-muegamma}.
This makes $\mu\to e\gamma$ a direct and stringent consistency condition on
the neutrino-mass solution. The MEG~II search using the 2021--2022 data found
no significant excess and established the most stringent upper limit to
date~\cite{MEGII:2025new},
\begin{equation}\label{eq:megii-limit}
 \mathcal{B}(\mu^+\to e^+\gamma)<1.5\times10^{-13}
 \qquad (90\%\ {\mathrm{C.L.}}).
\end{equation}

\begin{figure}[t!]
    \centering
    \includegraphics[width=\linewidth]{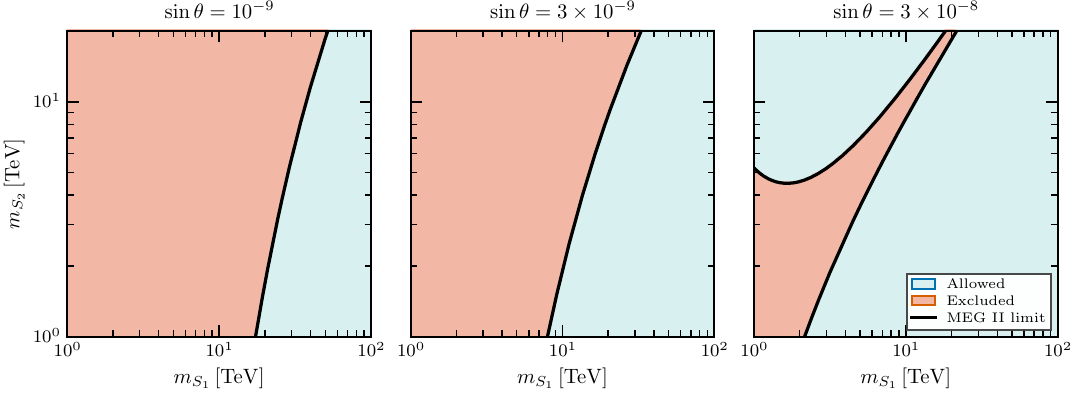}
    \caption{Constraint from the $Y^L$ contribution to $\mu\to e\gamma$ in the
    $(m_{S_1},m_{S_2})$ plane for $Y^R=\operatorname{diag}(1,1,1)$, using
    $\zeta=V_{\mathrm{CKM}}^{T}Y^L$, $m_\nu^D=\zeta^\dagger\mathcal A$, and the
    restricted normal-ordering texture $\zeta_{1\alpha}=0$. The panels show
    $\sin\theta=10^{-9}$ (left), $\sin\theta=3\times10^{-9}$ (center), and
    $\sin\theta=3\times10^{-8}$ (right). Blue regions satisfy the MEG~II bound
    in Eq.~\eqref{eq:megii-limit}, while orange regions are excluded. The black
    contours correspond to $\mathcal{B}(\mu\to e\gamma)=1.5\times10^{-13}$.}
    \label{fig:lfv-megii}
\end{figure}

Figure~\ref{fig:lfv-megii} shows the MEG~II constraint for the fixed benchmark
$Y^R=\operatorname{diag}(1,1,1)$. For $\sin\theta=10^{-9}$, the lower bound on
$m_{S_1}$ increases from approximately $17.4\,\mathrm{TeV}$ at
$m_{S_2}=1\,\mathrm{TeV}$ to $51.9\,\mathrm{TeV}$ at
$m_{S_2}=20\,\mathrm{TeV}$. For $\sin\theta=3\times10^{-9}$, these bounds
decrease to $8.0$ and $33.0\,\mathrm{TeV}$, respectively, showing that larger
scalar mixing weakens the CLFV constraint.

For $\sin\theta=3\times10^{-8}$, the exclusion is concentrated near
$m_{S_1}=m_{S_2}$, while either mass hierarchy is allowed sufficiently far
from degeneracy. This follows from Eqs.~\eqref{eq:alt-reconstruction}
and~\eqref{eq:alt-loop-function-heavy-limit}, since $Y^L$ scales inversely
with $\sin(2\theta)\ln(m_{S_1}^2/m_{S_2}^2)$. Scalar degeneracy therefore
enhances the dipole contribution and the sensitivity to the MEG~II limit.

We set $\widetilde Y^R=0$ in the CLFV analysis, defining a minimal benchmark
in which neutrino data fix the dipole amplitude once $Y^R$ and the
scalar-sector parameters are specified. The matrix $\widetilde Y^R$ introduces
independent flavor parameters that are not required for neutrino mass
generation or the tree-level $b\to s\nu\bar\nu$ transition. For nonzero
$\widetilde Y^R$, the amplitude also receives pure right-handed contributions proportional to
$\sum_i\widetilde Y^R_{i\mu}\widetilde Y^{R*}_{ie}$ and mixed-chirality terms
involving $Y^L$ and $\widetilde Y^R$. The latter involve an internal-quark
chirality flip, are weighted by the up-type-quark mass, and can interfere with
contributions to the same dipole chirality. A diagonal $\widetilde Y^R$
eliminates the pure right-handed flavor-changing combination, but not the
mixed terms because $Y^L$ is non-diagonal.

\section{$\Delta N_{\mathrm{eff}}$}\label{sec:neff}

We study $\Delta N_{\mathrm{eff}}\equiv N_{\mathrm{eff}}-N_{\mathrm{eff}}^{\mathrm{SM}}$,
the additional contribution to the effective number of relativistic
neutrino species. In this model, the $Y^R$ interactions can populate the
right-handed neutrinos in the early Universe. If they thermalize and later
decouple, their temperature is diluted relative to that of the active
neutrinos, yielding a contribution controlled by their decoupling
temperatures~\cite{Abazajian:2019oqj,Luo:2020fdt}.

For the cosmological constraint, we use the Planck 2018
TT,TE,EE+lowE+lensing+BAO result in the $\Lambda$CDM+$N_{\mathrm{eff}}$
extension, $N_{\mathrm{eff}}=2.99^{+0.34}_{-0.33}$ at $95\%$ confidence
level~\cite{Planck:2018parameters}. Together with the most recent SM prediction
$N_{\mathrm{eff}}^{\mathrm{SM}}=3.0440\pm0.0002$~\cite{Bennett:2020zkv, Mangano:2005cc}, this
motivates the approximate upper cut $\Delta N_{\mathrm{eff}}\leq0.286$ used
below. More recent results combining ACT DR6, Planck, and DESI Year-1 data
find $N_{\mathrm{eff}}=2.86\pm0.13$ at $68\%$ confidence
level~\cite{ACT:2025neff}; we retain Planck 2018+BAO as a conservative
reference and to facilitate comparison with previous studies.

\begin{figure}[t!]
\centering
\begin{tikzpicture}[every node/.style={font=\small}]
  \begin{feynman}
    \vertex (nu) at (-2.3,1) {$\nu_{iR}$};
    \vertex (nubar) at (-2.3,-1) {$\bar\nu_{jR}$};
    \vertex[dot, minimum size=3pt] (vT) at (0,1) {};
    \vertex[dot, minimum size=3pt] (vB) at (0,-1) {};
    \vertex (dbar) at (2.3,1) {$\bar d_i$};
    \vertex (d) at (2.3,-1) {$d_j$};
    \diagram*{
      (nu) -- [fermion] (vT),
      (nubar) -- [anti fermion] (vB),
      (vT) -- [scalar, edge label={$S_a$}] (vB),
      (vT) -- [anti fermion] (dbar),
      (vB) -- [fermion] (d)
    };
  \end{feynman}
\end{tikzpicture}
\caption{$t$-channel annihilation $\nu_{iR}\bar\nu_{jR}\to\bar d_i d_j$
through $S_a$ ($a=1,2$). Its charge-conjugate channel gives
$d_i\bar d_j$; the inverse reactions populate the right-handed neutrinos.}
\label{fig:neff-annihilation}
\end{figure}
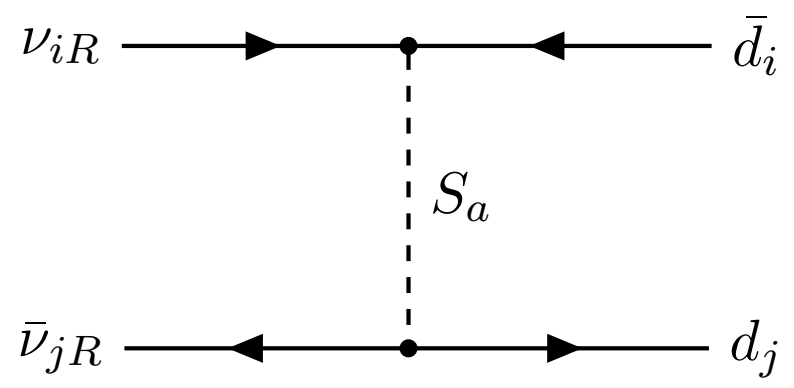

To determine whether the right-handed neutrinos thermalize, we consider the
$t$-channel annihilation $\nu_{iR}\bar\nu_{jR}\to\bar d_i d_j$, with
$d_{1,2,3}=d,s,b$, shown in Fig.~\ref{fig:neff-annihilation}. The inverse
reactions populate the right-handed states.\footnote{Crossed scattering
$\nu_{iR}d_i\leftrightarrow\nu_{jR}d_j$, including $i=j$, is not included.}

We neglect neutrino masses in the reaction kinematics while retaining the
down-quark masses and work at the free-quark level.\footnote{Below the QCD
transition, a quantitative treatment requires hadronic processes.} In terms
of the Mandelstam variables $s$ and $t$, the cross section for fixed initial
helicities, summed over final spins and colors, is
\begin{equation}
    \sigma_{ij}(s)=\frac{3|Y^R_{ii}Y^R_{jj}|^2}{16\pi s^2}
    \left[
    \sin^4\theta\,f(m_{S_1}^2)
    +\cos^4\theta\,f(m_{S_2}^2)
    +2\sin^2\theta\cos^2\theta\,I_{12}
    \right].
\label{eq:neff-cross-section}
\end{equation}
The function associated with each diagonal scalar contribution is
\begin{equation}
\begin{split}
    f(m_{S_a}^2)={}&(t_+-t_-)
    +(2m_{S_a}^2-m_{d_i}^2-m_{d_j}^2)
    \ln\!\left(\frac{t_+-m_{S_a}^2}{t_--m_{S_a}^2}\right) \\
    &+(m_{S_a}^2-m_{d_i}^2)(m_{S_a}^2-m_{d_j}^2)
    \left(\frac{1}{t_--m_{S_a}^2}-\frac{1}{t_+-m_{S_a}^2}\right).
\end{split}
\label{eq:neff-diagonal-function}
\end{equation}
The corresponding interference function is
\begin{equation}
    I_{12}=(t_+-t_-)
    +A\ln\!\left(\frac{t_+-m_{S_1}^2}{t_--m_{S_1}^2}\right)
    +B\ln\!\left(\frac{t_+-m_{S_2}^2}{t_--m_{S_2}^2}\right).
\label{eq:neff-interference-function}
\end{equation}
The kinematic quantities entering these functions are
\begin{equation}
\begin{aligned}
    t_{\pm}&=\frac{m_{d_i}^2+m_{d_j}^2-s
    \pm\sqrt{\lambda(s,m_{d_i}^2,m_{d_j}^2)}}{2},\\
    \lambda(x,y,z)&=x^2+y^2+z^2-2xy-2xz-2yz,
\end{aligned}
\label{eq:neff-kinematics}
\end{equation}
where $\lambda$ is the K\"all\'en function. Finally, the coefficients in
Eq.~\eqref{eq:neff-interference-function} are
\begin{equation}
\begin{aligned}
    A={}&\frac{(m_{S_1}^2-m_{d_i}^2)(m_{S_1}^2-m_{d_j}^2)}
    {m_{S_1}^2-m_{S_2}^2},
    \qquad
    B=\frac{(m_{S_2}^2-m_{d_i}^2)(m_{S_2}^2-m_{d_j}^2)}
    {m_{S_2}^2-m_{S_1}^2}.
\end{aligned}
\label{eq:neff-interference-coefficients}
\end{equation}
The cross section vanishes below the threshold
$s=(m_{d_i}+m_{d_j})^2$.\footnote{Although $A$ and $B$ are separately
singular for equal scalar masses, their combination is finite:
$I_{12}\to f(M^2)$ as $m_{S_1}^2,m_{S_2}^2\to M^2$.}

Let $n_{\nu_R}^{\mathrm{eq}}(T)$ denote the equilibrium density of one generation,
including both $\nu_{iR}$ and $\bar\nu_{iR}$. At a common temperature $T$ and
vanishing chemical potentials, this density is the same for all generations:
\begin{equation}
    n_{\nu_R}^{\mathrm{eq}}(T)=\frac{3}{4}\frac{\zeta(3)}{\pi^2}
    g_{\nu_R}T^3.
\label{eq:neff-number-density}
\end{equation}
Here, $\zeta$ is the Riemann zeta function and $g_{\nu_R}=2$ counts one
helicity state for the neutrino and one for its antiparticle. Consequently,
$n_{\nu_{iR}}^{\mathrm{eq}}=n_{\bar\nu_{iR}}^{\mathrm{eq}}=n_{\nu_R}^{\mathrm{eq}}/2$.

Using Fermi--Dirac distributions for both incoming particles and the
M{\o}ller flux factor $v_{\rm M}$ in the plasma rest frame~\cite{Cannoni:2013bza, Gondolo:1990dk},
we write the thermal average as
\begin{equation}
\begin{split}
    \langle\sigma_{ij}v_{\mathrm{M}}\rangle
    ={}&\frac{4}{[n_{\nu_R}^{\mathrm{eq}}(T)]^2}
    \int\frac{d^3p_1}{(2\pi)^3}\frac{d^3p_2}{(2\pi)^3}\\
    &\times f_{\mathrm{FD}}(E_1,T)f_{\mathrm{FD}}(E_2,T)\,
    \sigma_{ij}(s)v_{\mathrm{M}}.
\end{split}
\label{eq:neff-thermal-average}
\end{equation}
Here, $f_{\mathrm{FD}}(E,T)=(e^{E/T}+1)^{-1}$,
$E_a=|\mathbf p_a|$, and
$v_{\mathrm{M}}=1-\widehat{\mathbf p}_1\cdot\widehat{\mathbf p}_2$.
The momenta $p_1$ and $p_2$ correspond to the incoming neutrino and
antineutrino, respectively.
We neglect final-state Pauli blocking, so that the vacuum cross section
in Eq.~\eqref{eq:neff-cross-section} can be used.

The annihilation rate per $\nu_{iR}$ is obtained by summing over the
antineutrino species,
\begin{equation}
    \Gamma_i(T)=\frac{n_{\nu_R}^{\mathrm{eq}}(T)}{2}\sum_j
    \langle\sigma_{ij}v_{\mathrm{M}}\rangle.
\label{eq:neff-interaction-rate}
\end{equation}
We compare this rate with the expansion rate during radiation domination,
where $g_*(T)$ counts the effective energy degrees of freedom,
\begin{equation}
    H(T)=1.66\sqrt{g_*(T)}\frac{T^2}{M_{\mathrm{Pl}}}.
\label{eq:neff-hubble-rate}
\end{equation}
Here $M_{\mathrm{Pl}}=G_N^{-1/2}\simeq1.22\times10^{19}\,\mathrm{GeV}$
is the unreduced Planck mass, with $G_N$ the gravitational constant.
For both $g_*(T)$ and the entropy degrees of freedom $g_{*s}(T)$ below,
we use Table~A1 of Ref.~\cite{Husdal:2016haj}, with its
$T_{\mathrm{QCD}}=214\,\mathrm{MeV}$ prescription. We interpolate linearly
in $\ln T$ within each phase, retaining the tabulated entropy jump
at the transition; both functions approach $106.75$ at high temperature.

For species that reach thermal equilibrium, we use the instantaneous
decoupling approximation and compute the subsequent entropy dilution
following Ref.~\cite{Anchordoqui:2012qu}. We estimate the decoupling
temperature of $\nu_{iR}$ from
\begin{equation}
    \Gamma_i(T_{\mathrm{dec},i})=H(T_{\mathrm{dec},i}).
\label{eq:neff-decoupling}
\end{equation}
We select the low-temperature crossing at which $\Gamma_i/H$ falls below
unity as the Universe cools. Because the full cross section depends
nontrivially on $s$, this equation is solved numerically.
We assume a negligible initial abundance. For species that never reach
thermal equilibrium over the assumed thermal history, we neglect
production by freeze-in~\cite{Luo:2020fdt}.
The entropy-dilution formula below applies to the species that did
thermalize.

\begin{figure}[t!]
    \centering
    \includegraphics[width=0.78\linewidth]{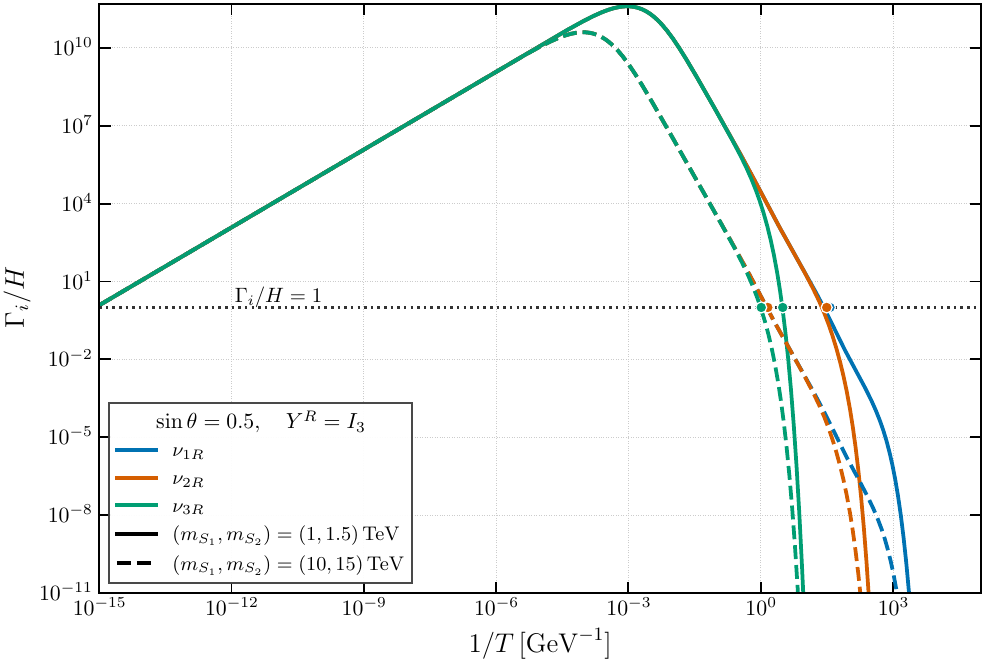}
    \caption{Annihilation rates $\Gamma_i/H$ as functions of inverse
    temperature for $Y^R=\operatorname{diag}(1,1,1)$ and $\sin\theta=0.5$.
    Solid curves correspond to
    $(m_{S_1},m_{S_2})=(1,1.5)\,\mathrm{TeV}$ and dashed curves to
    $(10,15)\,\mathrm{TeV}$. Markers identify the decoupling crossings
    $\Gamma_i/H=1$.}
    \label{thermal-equilibrium}
\end{figure}

\begin{figure}[t!]
    \centering
    \includegraphics[width=0.78\linewidth]{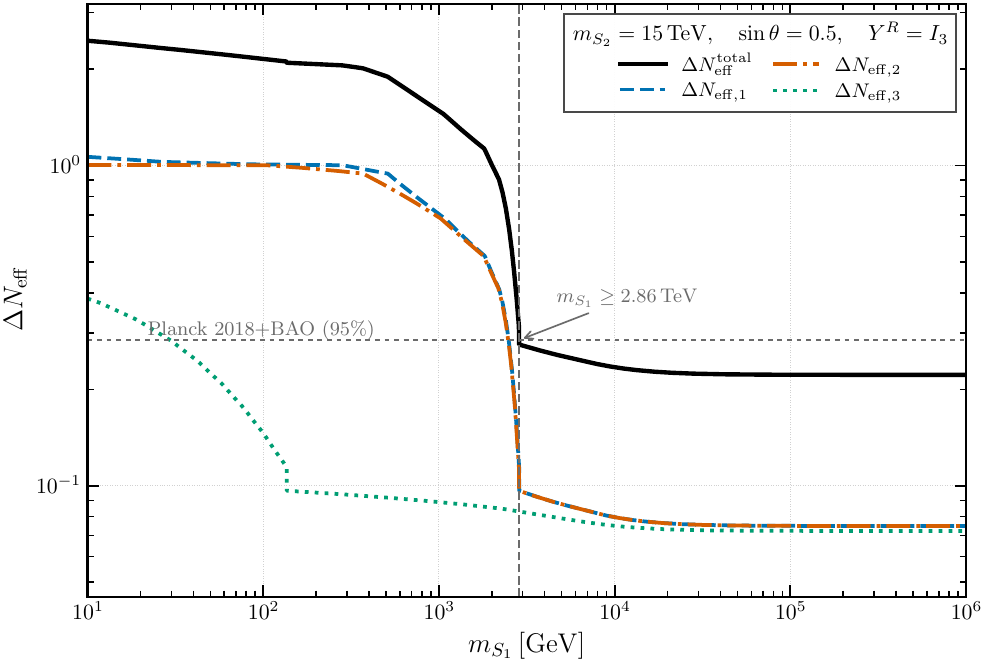}
    \caption{Individual and total $\Delta N_{\mathrm{eff}}$ contributions as
    functions of $m_{S_1}$ for $m_{S_2}=15\,\mathrm{TeV}$,
    $\sin\theta=0.5$, and $Y^R=\operatorname{diag}(1,1,1)$. The horizontal
    dotted line shows the adopted Planck 2018+BAO $95\%$ bound; the vertical
    dashed line marks the resulting lower bound on $m_{S_1}$.}
    \label{fig:delta_neff}
\end{figure}

The tree-level thermal rates depend only on $Y^R$, the scalar masses,
and the mixing angle $\theta$. Figure~\ref{thermal-equilibrium} illustrates
the resulting evolution of $\Gamma_i/H$ for two representative scalar
spectra; the relevant decoupling temperatures are the low-temperature
crossings of unity.
Entropy conservation after instantaneous decoupling fixes the temperature
of each right-handed species relative to that of the active neutrinos:
\begin{equation}
    \frac{T_{\nu_{iR}}}{T_{\nu_L}}
    =\left[
    \frac{g_{*s}(T_{\mathrm{dec}}^{\nu_L})}
    {g_{*s}(T_{\mathrm{dec},i})}
    \right]^{1/3}.
\label{eq:neff-temperature-ratio}
\end{equation}
Here $T_{\mathrm{dec}}^{\nu_L}$ denotes active-neutrino decoupling. Taking
$g_{*s}(T_{\mathrm{dec}}^{\nu_L})=10.75$, before $e^\pm$ annihilation, the
$\nu_{iR},\bar\nu_{iR}$ pair contributes
\begin{equation}
    \Delta N_{\mathrm{eff},i}
    =\left(\frac{T_{\nu_{iR}}}{T_{\nu_L}}\right)^4
    =\left[
    \frac{10.75}{g_{*s}(T_{\mathrm{dec},i})}
    \right]^{4/3}.
\label{eq:neff-single}
\end{equation}
Since the three generations can decouple at different temperatures, their
contributions add independently:
\begin{equation}\label{eq:neff-total}
    \Delta N_{\mathrm{eff}}
    =\sum_{i=1}^{3}\Delta N_{\mathrm{eff},i}.
\end{equation}
Figure~\ref{fig:delta_neff} shows the individual and total contributions for
a representative one-dimensional mass scan, together with the adopted
cosmological bound.

\begin{figure}[t!]
    \centering
    \includegraphics[width=0.68\linewidth]{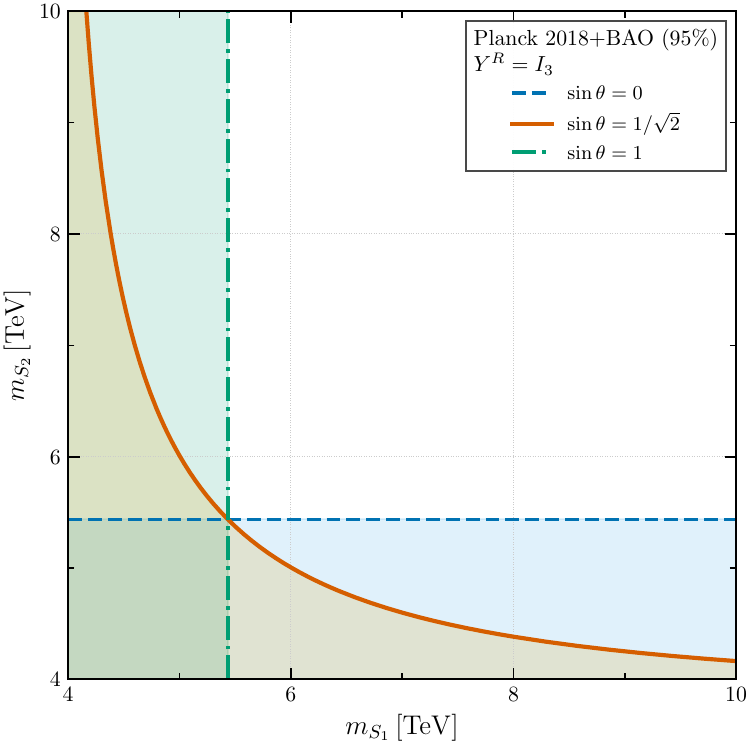}
\caption{Planck 2018+BAO $95\%$ exclusion regions in the
$(m_{S_1},m_{S_2})$ plane for $Y^R=\operatorname{diag}(1,1,1)$ and
$\sin\theta=0,\ 1/\sqrt{2},\ 1$. The curves mark the adopted
$\Delta N_{\mathrm{eff}}$ bound and shaded regions are excluded. At maximal
mixing, the contour is symmetric under $m_{S_1}\leftrightarrow m_{S_2}$.}
    \label{fig:exclusion_neff}
\end{figure}

\FloatBarrier
Figure~\ref{fig:exclusion_neff} translates the same bound into the scalar-mass
plane for three mixing choices. At maximal mixing, the equal scalar weights
produce an exclusion contour symmetric under
$m_{S_1}\leftrightarrow m_{S_2}$. The endpoint curves at $\sin\theta=0$ and
$1$ are thermal limits; since $\sin(2\theta)=0$, they do not generate
radiative neutrino masses.

For equal scalar masses at maximal mixing, Fig.~\ref{fig:exclusion_neff}
gives $m_{S_1}=m_{S_2}\gtrsim5.4\,\mathrm{TeV}$ for
$Y^R=I_3$. A direct calculation with the same thermal treatment and
$Y^R=\sqrt{4\pi}\,I_3$, near the perturbative limit, raises this lower bound
to approximately $19.3\,\mathrm{TeV}$. For this particular configuration,
the cosmological bound lies more than an order of magnitude above the
characteristic $\mathcal O(1\,\mathrm{TeV})$ limits from current direct
searches, illustrating the potential strength of $N_{\mathrm{eff}}$.
This comparison is indicative, since a quantitative LHC limit requires a
dedicated recast of the mixed-leptoquark
spectrum~\cite{ATLAS:2025LQsummary,ATLAS:2025resonant}.

\FloatBarrier
\section{Numerical analysis and discussion}
\label{sec:numerical-analysis}

We now combine the neutrino-mass, meson-decay, charged-lepton flavor,
and cosmological results to identify viable Yukawa textures.
Fig.~\ref{fig:numerical-cumulative} shows how the constraints accumulate
at fixed scalar spectra, while Fig.~\ref{fig:numerical-mass-map} displays
the resulting reach in the scalar mass plane.

\subsection{Physical inputs and constraints}

We adopt the restricted flavor texture and neutrino-mass reconstruction
defined in Eqs.~\eqref{eq:alt-numerical-texture}
and~\eqref{eq:alt-singular-texture}. Thus, at each scalar spectrum and
mixing angle, only the positive parameters $y_2=Y^R_{22}$ and
$y_3=Y^R_{33}$ are varied, while $\zeta$ and $Y^L=V^*\zeta$ are
reconstructed using the exact loop function in
Eq.~\eqref{eq:alt-loop-function}. Neutrino masses and mixing are therefore
imposed algebraically. The elementwise perturbativity condition in
Eq.~\eqref{eq:alt-perturbativity-condition} is applied to $Y^R$, $\zeta$,
and $Y^L$, retaining the CKM-induced correlations in
Eq.~\eqref{eq:alt-singular-texture}.

For reproducibility, we use the NuFIT~6.0 inputs collected in
Eq.~\eqref{eq:alt-oscillation-inputs}. The PMNS and CKM phases are retained in
Eq.~\eqref{eq:alt-singular-texture}; oscillation and form-factor uncertainties are not
scanned.

The objective function is based on the inclusive
$B^+\to K^+\nu\bar\nu$ branching fraction defined in
Eq.~\eqref{eq:BR}, with the differential rate given in
Eq.~\eqref{eq:general-rate}. We adopt the combined Belle~II measurement and its
$1\sigma$ interval~\cite{Belle-II:2023esi},
$[1.6596876,3.0071068]\times10^{-5}$, obtained by combining statistical
and systematic errors in quadrature on each side. The objective
vanishes inside this interval and penalizes the squared departure
outside it. Two additional requirements are imposed:
$\mathcal{B}(\mu\to e\gamma)\leq1.5\times10^{-13}$, from
Sec.~\ref{sec:clfv}, Eqs.~\eqref{eq:br-muegamma}
and~\eqref{eq:megii-limit}~\cite{MEGII:2025new}, and
$\Delta N_{\mathrm{eff}}\leq0.286$, evaluated according to
Sec.~\ref{sec:neff}, Eq.~\eqref{eq:neff-total}~\cite{Planck:2018parameters}.
These criteria use their respective experimental confidence levels
and do not define a joint confidence region.

Rare top decays can also be induced in this model but are not included among
the numerical constraints because their rates are negligible for the
$\widetilde Y^R=0$ benchmark adopted here; see
Appendix~\ref{sec:top-decays}. A nonzero $\widetilde Y^R$ would modify both
the CLFV and rare-top phenomenology and require a separate multidimensional
analysis.

Freeze-in production is not included. In particular, $\nu_{1R}$ has no
thermal contribution for the adopted $Y^R_{11}=0$ texture.

\subsection{Combined constraints and scalar mass reach}

\begin{figure}[t!]
 \centering
 \includegraphics[width=\linewidth]{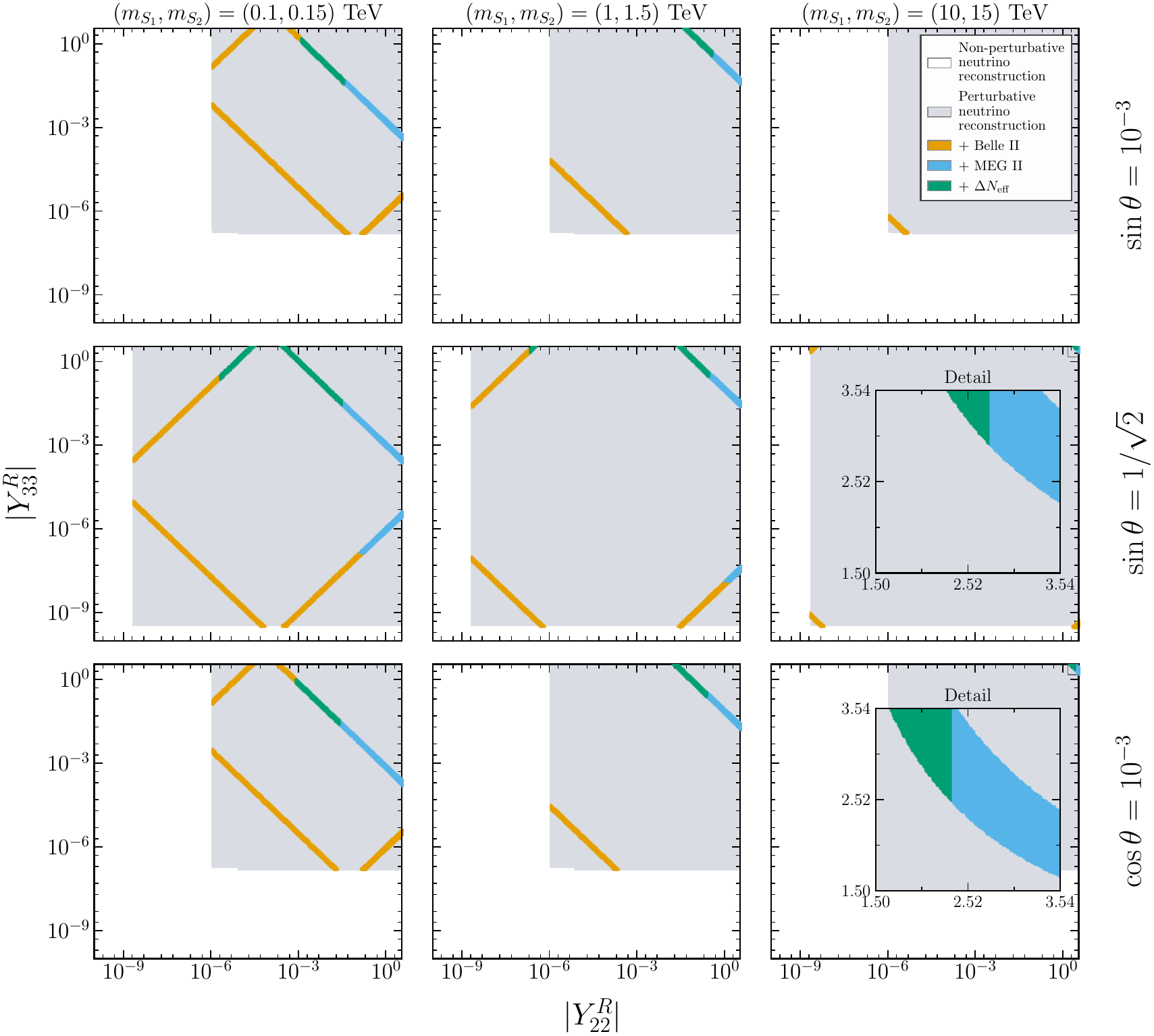}
 \caption{Cumulative constraints in the $(y_2,y_3)$ plane.
Columns show the indicated scalar spectra; rows correspond to
$\sin\theta=10^{-3}$, $\sin\theta=1/\sqrt{2}$, and
$\cos\theta=10^{-3}$. White corresponds to a nonperturbative neutrino
reconstruction; light gray has a perturbative neutrino reconstruction;
orange also satisfies Belle~II; blue additionally satisfies MEG~II; and
green satisfies all four conditions. The constraints are applied
cumulatively in the order shown in the legend.}
 \label{fig:numerical-cumulative}
\end{figure}

Fig.~\ref{fig:numerical-cumulative} shows the cumulative effect of
perturbativity, Belle~II, MEG~II, and $\Delta N_{\mathrm{eff}}$ for three scalar
spectra and mixing choices. For each fixed scalar spectrum and mixing angle,
we evaluate a $201\times201$ logarithmic grid in $(y_2,y_3)$, supplemented by
401 points resolving the Belle~II contours. The four conditions are then
applied cumulatively at each sampled point.

The Belle~II bands contain two regimes. For larger $y_2$ and $y_3$, the
reconstructed $\zeta$ and $Y^L$ entries are suppressed, and the RR vector
contribution dominates. A strong hierarchy between $y_2$ and $y_3$ instead
enhances one row of $\zeta$ and can produce a mixed-chirality scalar--tensor
solution.\footnote{Here ``RR vector'' denotes the contribution associated
with $C^V_{RR}$, generated by two right-handed vertices, while
``mixed-chirality scalar--tensor'' denotes the scalar and tensor operators
generated by vertices of opposite chirality; see
Appendix~\ref{app:meson-decay-details}.} MEG~II preferentially removes the latter regime, whereas
$\Delta N_{\mathrm{eff}}$ constrains larger right-handed Yukawas. Both mechanisms
survive for the $(1,1.5)\,\mathrm{TeV}$ spectrum at maximal mixing.
As the mixing approaches its endpoints, however, the radiative mass relation
requires larger Yukawa products, while the scalar weights increasingly select
different mass eigenstates.
For $(m_{S_1},m_{S_2})=(10,15)\,\mathrm{TeV}$, no sampled point survives MEG~II at
$\sin\theta=10^{-3}$, while RR-vector solutions remain at maximal mixing
and for $\cos\theta=10^{-3}$.

\begin{figure}[t!]
 \centering
 \includegraphics[width=\linewidth]{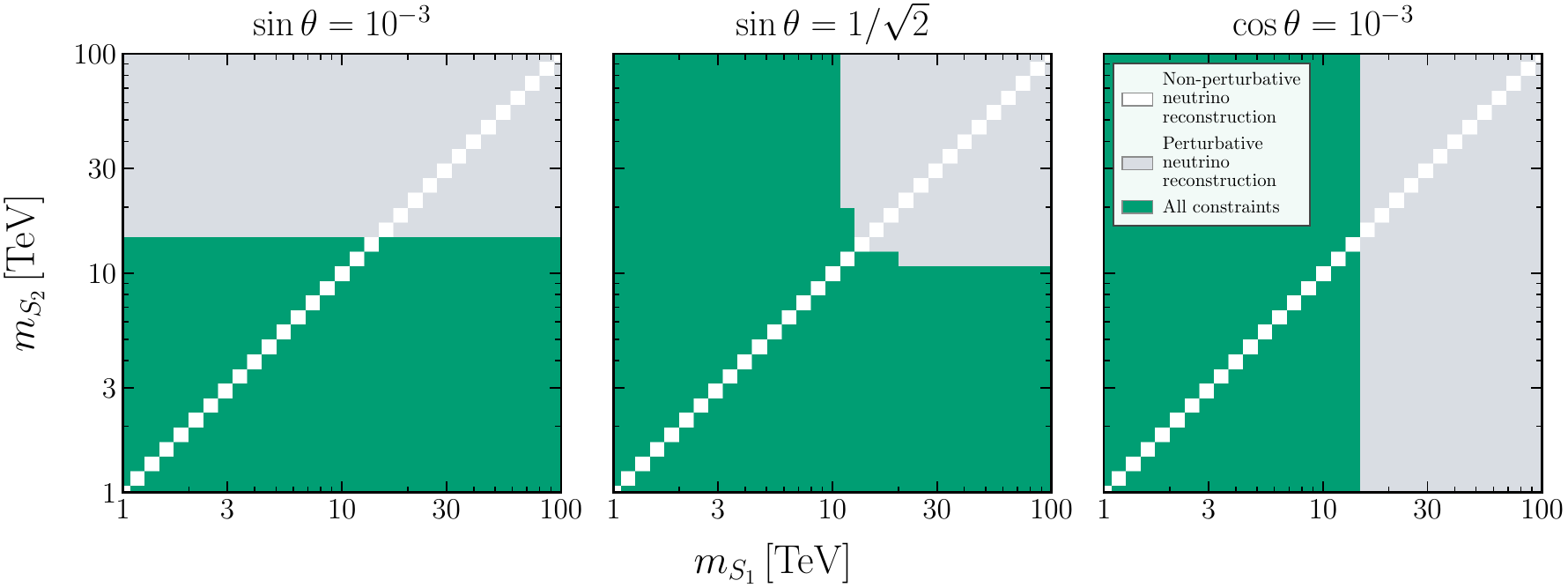}
 \caption{Search for textures satisfying all four conditions on a
$31\times31$ logarithmic mass grid from $1$ to $100\,\mathrm{TeV}$.
At each mass pair, white indicates that no perturbative neutrino-mass
texture can be reconstructed, light gray that a perturbative texture can
be reconstructed but no texture satisfying all constraints was found,
and green that at least one texture satisfying all constraints was found.}
 \label{fig:numerical-mass-map}
\end{figure}

Fig.~\ref{fig:numerical-mass-map} extends the search to independently
varied scalar masses. On a $31\times31$ logarithmic mass grid, we scan the
positive Yukawa parameters $y_2,y_3>0$ for each mass pair and mixing angle,
refining the search near Belle~II-compatible regions. As both
scalar masses increase, propagator suppression drives the branching fraction
back toward its SM value, making it too small to reproduce the Belle~II
measurement with perturbative Yukawa couplings. Consequently, no complete
solution is found once both masses exceed approximately
$12$--$14\,\mathrm{TeV}$, although one mass can reach $100\,\mathrm{TeV}$ if
the other remains light. The endpoint panels are approximately transposed
because the scalar weights interchange. The lower explored mass of
$1\,\mathrm{TeV}$ is motivated by current collider bounds on this scale,
although a dedicated recast is required for the present two-leptoquark
framework~\cite{ATLAS:2025LQsummary,ATLAS:2025resonant}.

Table~\ref{tab:numerical-benchmarks} makes the interplay among the
constraints explicit through four maximal-mixing textures selected from
Fig.~\ref{fig:numerical-cumulative}. Benchmark A is RR-vector dominated and
has negligible flavor violation, whereas B realizes the mixed-chirality
regime, with larger left-handed couplings and a $\mu\to e\gamma$ rate near
the MEG~II bound. Benchmark C is again RR-vector dominated but instead lies
close to the cosmological limit. Finally, D accommodates the Belle~II
enhancement at $(m_{S_1},m_{S_2})=(10,15)\,\mathrm{TeV}$ at the cost of
right-handed Yukawa couplings of order three. All four reproduce the same
neutrino spectrum and mixing parameters, while their meson, flavor, and
cosmological predictions highlight the complementarity of the constraints.
We emphasize that, for each benchmark, $Y^R$ is diagonal, with the two
positive nonzero entries listed in Table~\ref{tab:numerical-benchmarks}; the
matrices $\zeta$ and $Y^L=V^*\zeta$ then follow uniquely from the
reconstruction in Eq.~\eqref{eq:alt-singular-texture}.

\begin{table}[htbp]
\centering
\small
\renewcommand{\arraystretch}{1.2}
\setlength{\tabcolsep}{8.1pt}
\begin{tabular}{l c c c c}
\hline\hline
Parameter or observable & A & B & C & D \\ \hline
$m_{S_1}$ [$\mathrm{TeV}$] & 1 & 1 & 1 & 10 \\
$m_{S_2}$ [$\mathrm{TeV}$] & 1.5 & 1.5 & 1.5 & 15 \\
$\sin\theta$ & $1/\sqrt{2}$ & $1/\sqrt{2}$ & $1/\sqrt{2}$ & $1/\sqrt{2}$ \\
$|Y^R_{22}|$ & $0.0981$ & $2.102\times10^{-7}$ & $0.2747$ & $2.7119$ \\
$|Y^R_{33}|$ & $1.0300$ & $1.9558$ & $0.4296$ & $3.2445$ \\
$\max_{ij}|Y^L_{ij}|$ & $7.812\times10^{-8}$ & $0.0365$ & $2.799\times10^{-8}$ & $2.839\times10^{-9}$ \\
$10^5\mathcal{B}(B^+\to K^+\nu\bar\nu)$ & 2.3334 & 1.6597 & 3.0071 & 1.8843 \\
$\mathcal{B}(\mu\to e\gamma)$ & $3.038\times10^{-36}$ & $1.444\times10^{-13}$ & $4.930\times10^{-38}$ & $5.180\times10^{-46}$ \\
$\Delta N_{\mathrm{eff}}$ & 0.1708 & 0.0870 & 0.2785 & 0.2606 \\
RR vector share [\%] & 79.43 & $<10^{-8}$ & 84.04 & 74.53 \\
Scalar + tensor share [\%] & $<10^{-8}$ & 71.09 & $<10^{-8}$ & $<10^{-8}$ \\ \hline\hline
\end{tabular}
\caption{Four illustrative textures from the maximal-mixing panels of Fig.~\ref{fig:numerical-cumulative}, with $\widetilde Y^R=0$. All satisfy the adopted Belle~II, MEG~II, and cosmological conditions and the perturbativity criterion in Eq.~\eqref{eq:alt-perturbativity-condition}. The neutrino masses are the same for all four: $m_1=0$, $m_2=8.654\times10^{-3}\,\mathrm{eV}$, and $m_3=5.013\times10^{-2}\,\mathrm{eV}$, with the oscillation inputs specified in the text. Operator shares refer to the total meson branching fraction in Eq.~\eqref{eq:general-rate}, including the SM contribution. Values are rounded; acceptance is checked with the stored full-precision Yukawa matrices. Collider limits are not imposed in this table.}
\label{tab:numerical-benchmarks}
\end{table}

\section{Conclusions}
\label{sec:conclusions}

We have studied a simple leptoquark model connecting radiative Dirac
neutrino masses with the enhancement of $B^+\to K^+\nu\bar\nu$.
The scalar mixing needed for the one-loop mass mechanism also links
the left- and right-handed interactions entering the meson decay.
The renormalizable model conserves baryon and lepton numbers, preserving
proton stability and the Dirac nature of neutrinos after soft breaking
of the discrete symmetries.

Within the adopted flavor ansatz, the neutrino-mass reconstruction
leaves two right-handed Yukawa entries free. Their allowed values are
restricted jointly by perturbativity, the Belle~II branching fraction
interval, $\mu\to e\gamma$, and $\Delta N_{\mathrm{eff}}$.
Figs.~\ref{fig:numerical-cumulative} and~\ref{fig:numerical-mass-map}
show that compatible textures occur in both the RR-vector-dominated and
mixed-chirality scalar--tensor regimes. The latter can approach
the charged-lepton flavor bound through larger reconstructed
left-handed couplings, while RR-vector solutions can approach the
cosmological bound with negligible flavor violation. We provide four
benchmarks that make this complementarity explicit. Under the baseline
assumption
$\widetilde Y^R=0$, rare top decays remain suppressed for the flavor structure
considered and do not supply an additional restriction in the numerical study.

Within the adopted texture and scan, no complete solution is found once both
scalar masses exceed approximately $12$--$14\,\mathrm{TeV}$, although one
scalar can be much heavier if the other remains sufficiently light. Direct
collider searches provide the complementary restriction at low masses, but
existing LHC limits cannot be applied quantitatively without a dedicated
recast. The two mixed scalar leptoquarks originate from fields with identical
gauge quantum numbers but distinct Yukawa structures, leading to
flavor-dependent decay patterns~\cite{ATLAS:2022wcu,ATLAS:2025LQsummary,ATLAS:2025resonant}.
Consequently, the lower mass reach is not universal. Nevertheless,
$\Delta N_{\mathrm{eff}}$ provides a powerful probe of leptoquark scenarios
with radiative Dirac neutrino masses and, for some benchmark configurations,
can surpass the nominal reach of current direct searches.

\section*{Acknowledgments}
P. E. acknowledges support from CNPq under grant No.~151612/2024-2 and from
the Fundação de Amparo e Promoção da Ciência, Tecnologia e Inovação do Rio
Grande do Norte (FAPERN) under Call No.~06/2026--FAPERN/IIP, process
No.~10910022.000493/2024-04. G. G. acknowledges support from the National
Council for Science, Technology and Technological Innovation (CONCYTEC) and
the National Program for Scientific Research and Advanced Studies
(PROCIENCIA) under the call ``E041-2026-01 Proyectos de Investigación
Básica'' (contract No.~PE5010103523-2026), and thanks the International
Institute of Physics (IIP-UFRN) in Natal for its kind hospitality, financial
support, and stimulating research environment during the development of this
work. V. P. acknowledges financial support from the Conselho Nacional de
Desenvolvimento Científico e Tecnológico (CNPq). This study was financed in
part by the Coordenação de Aperfeiçoamento de Pessoal de Nível Superior --
Brasil (CAPES) -- Finance Code 001. C. A. S. P. thanks CNPq for financial
support under grant No.~301734/2026-7. F. S. Q. is supported by the Humboldt
Foundation (Friedrich Wilhelm Bessel Research Award), ANID -- Millennium
Science Initiative Program -- ICN2019\_044, IIF-FINEP grant 213/2024, FAPESP
grants 2021/14335-0 and 2021/00449-4, and CNPq grants 307130/2021-5,
403521/2024-6, 309465/2026-5, and 446711/2025-0.

\appendix

\section{Meson decay details}
\label{app:meson-decay-details}

\subsection{Differential rate and tree-level matching}
In this appendix, we detail the calculation of the inclusive branching
fraction shown in Eq.~\eqref{eq:BR} of Sec.~\ref{sec:meson-decays}.
Neglecting neutrino masses in the decay kinematics, the differential rate for the dimension-six vector, scalar, and tensor operators is
given by~\cite{Browder:2021hbl}
\begin{align}
\frac{d\Gamma}{d q^2}=&
\frac{G_F^2|V_{tb}V_{ts}^*|^2\alpha_{\mathrm{em}}^2}
{3072\pi^5m_B^3}
q^2\lambda^{1/2}_K(q^2) \nonumber \\
&\times \sum_{i,j=1}^3\Bigg\{
\Big(
\left|C_{LL}^{\mathrm{SM}}\delta_{ij}+\left[C^V_{LL}\right]_{ij}+\left[C^V_{RL}\right]_{ij}\right|^2
\nonumber
+\left| \left[C^V_{LR}\right]_{ij}+\left[C^V_{RR}\right]_{ij}\right|^2
\Big)H_V^2
\nonumber\\
&+\frac32\Big(
\left| \left[C^S_{RL}\right]_{ij}+\left[C^S_{LL}\right]_{ij}\right|^2
+\left|\left[C^S_{RR}\right]_{ij}+\left[C^S_{LR}\right]_{ij}\right|^2
\Big)H_S^2
\nonumber\\
&+8\Big(
\left|\left[C^T_{LL}\right]_{ij}\right|^2+\left|\left[C^T_{RR}\right]_{ij}\right|^2
\Big)H_T^2
\Bigg\},
\label{eq:general-rate}
\end{align}
where $G_F$ is the Fermi constant, $\alpha_{\mathrm{em}}$ is the
electromagnetic fine-structure constant, $m_B$ is the charged-$B$-meson
mass, and $V_{qq'}$ are CKM matrix elements.
The dimensionless SM Wilson coefficient is
$C_{LL}^{\mathrm{SM}}=-2X_t/\sin^2\theta_W$, for which we use
$X_t=1.469$, $\sin^2\theta_W=0.23121$, and
$\alpha_{\mathrm{em}}^{-1}(M_Z)=127.952$~\cite{Parrott:2022sm}. The CKM elements
are evaluated using the standard-parametrization inputs specified in
Sec.~\ref{sec:framework}.

At tree level, the nonzero leptoquark coefficients are
$C^{V,S,T}_{LL,a}$ and $C^{V,S,T}_{RR,a}$ for each physical scalar
$S_a$ ($a=1,2$). At the matching scale $\mu=m_{S_a}$ they read
\begin{align}
\big[C^{V}_{LL,a}(m_{S_a})\big]_{ij}
&=
\mathcal{K_B}\,
\frac{
\big(\mathcal{Y}^{(a)}_{L}\big)_{bj}
\big(\mathcal{Y}^{(a)}_{L}\big)^{*}_{si}
}
{m_{S_a}^{2}},
&
\big[C^{V}_{RR,a}(m_{S_a})\big]_{ij}
&=
\mathcal{K_B}\,
\frac{
\big(\mathcal{Y}^{(a)}_{R}\big)_{bj}
\big(\mathcal{Y}^{(a)}_{R}\big)^{*}_{si}
}
{m_{S_a}^{2}},
\label{eq:table-C-vector}
\end{align}

\begin{align}
\big[C^{S}_{LL,a}(m_{S_a})\big]_{ij}
&=
\mathcal{K_B}\,
\frac{
\big(\mathcal{Y}^{(a)}_{L}\big)_{bj}
\big(\mathcal{Y}^{(a)}_{R}\big)^{*}_{si}
}
{m_{S_a}^{2}},
&
\big[C^{S}_{RR,a}(m_{S_a})\big]_{ij}
&=
\mathcal{K_B}\,
\frac{
\big(\mathcal{Y}^{(a)}_{R}\big)_{bj}
\big(\mathcal{Y}^{(a)}_{L}\big)^{*}_{si}
}
{m_{S_a}^{2}},
\label{eq:table-C-mixed-LL}
\end{align}
\begin{align}
\big[C^{T}_{LL,a}(m_{S_a})\big]_{ij}
&=
-\frac{1}{4}
\big[C^{S}_{LL,a}(m_{S_a})\big]_{ij},
&
\big[C^{T}_{RR,a}(m_{S_a})\big]_{ij}
&=
-\frac{1}{4}
\big[C^{S}_{RR,a}(m_{S_a})\big]_{ij},
\label{eq:table-C-mixed-RR}
\end{align}
where $\mathcal{K_B}\equiv \pi v^2/(\alpha_{\mathrm{em}}V_{tb}V_{ts}^{*})$
and the physical Yukawa couplings $\mathcal{Y}_{L,R}^{(a)}$ are defined
in Eq.~\eqref{eq:meson-physical-yukawa}.

\subsection{QCD evolution}
The Wilson coefficients entering Eq.~\eqref{eq:general-rate} must be
evaluated at a common low-energy scale. We choose
$\mu_b=4.8\,\mathrm{GeV}$, of the order of the bottom-quark mass and equal
to the scale at which the tensor form factor
$f_T(q^2,\mu)$ is provided in Ref.~\cite{Parrott:2022rgu}. This choice avoids
an additional evolution of $f_T$. Accordingly, the scalar and tensor
Wilson coefficients and the running quark masses entering the scalar
matrix element are also evaluated at $\mu_b=4.8\,\mathrm{GeV}$.
Since the two leptoquarks can have different masses, each contribution
must first be evolved from its own matching scale and only then added:
\begin{equation}
\big[C^X_{AA}(\mu_b)\big]_{ij}
=
\sum_{a=1}^{2}
\Omega_X(\mu_b,m_{S_a})
\big[C^X_{AA,a}(m_{S_a})\big]_{ij},
\qquad X=V,S,T,\quad A=L,R.
\label{eq:WC-QCD-running}
\end{equation}
At leading-logarithmic order, for $m_{S_a}>m_t$, the QCD evolution
factor is~\cite{Browder:2021hbl}
\begin{equation}
\Omega_X(\mu_b,m_{S_a})
=
\left[
\frac{\alpha_s(\mu_b)}{\alpha_s(m_t)}
\right]^{-\gamma_X/\beta_1^{(5)}}
\left[
\frac{\alpha_s(m_t)}{\alpha_s(m_{S_a})}
\right]^{-\gamma_X/\beta_1^{(6)}},
\qquad
\beta_1^{(n_f)}=\frac{2n_f-33}{6},
\label{eq:QCD-running-factor}
\end{equation}
where $n_f$ denotes the number of active quark flavors at the
corresponding renormalization scale, and
\begin{equation}
\gamma_V=0,
\qquad
\gamma_S=2,
\qquad
\gamma_T=-\frac{2}{3}.
\label{eq:QCD-anomalous-dimensions}
\end{equation}
Consequently, the vector coefficients do not run,
$\Omega_V=1$, whereas the scalar and tensor coefficients evolve as
\begin{align}
\Omega_S(\mu_b,M)
&=
\left[
\frac{\alpha_s(\mu_b)}{\alpha_s(m_t)}
\right]^{12/23}
\left[
\frac{\alpha_s(m_t)}{\alpha_s(M)}
\right]^{4/7},
\nonumber\\
\Omega_T(\mu_b,M)
&=
\left[
\frac{\alpha_s(\mu_b)}{\alpha_s(m_t)}
\right]^{-4/23}
\left[
\frac{\alpha_s(m_t)}{\alpha_s(M)}
\right]^{-4/21}.
\label{eq:scalar-tensor-QCD-factors}
\end{align}
The matching relation
$C^T_{AA,a}(m_{S_a})=-C^S_{AA,a}(m_{S_a})/4$
holds at each leptoquark matching scale. It does not hold at $\mu_b$,
because the scalar and tensor operators have different QCD anomalous
dimensions.

\subsection{Helicity amplitudes and form factors}
Neutrino masses are retained in the reconstruction of $\zeta$ but neglected
in the decay kinematics. The Källén function is
\begin{equation}
 \lambda_K(q^2)=m_B^4+m_K^4+(q^2)^2
 -2m_B^2m_K^2-2m_K^2q^2-2m_B^2q^2,
\label{eq:kallen}
\end{equation}
and
\begin{align}
H_V(q^2)&=\sqrt{\frac{\lambda_K(q^2)}{q^2}}f_+(q^2),\nonumber\\
H_S(q^2)&=\frac{m_B^2-m_K^2}{m_b(\mu_b)-m_s(\mu_b)}f_0(q^2),\nonumber\\
H_T(q^2)&=-\frac{\sqrt{\lambda_K(q^2)}}{m_B+m_K}f_T(q^2,\mu_b).
\label{eq:helicity}
\end{align}
The lattice-QCD results for the form factors in
Ref.~\cite{Parrott:2022rgu} are fitted using the
Bourrely--Caprini--Lellouch (BCL) parametrization~\cite{BCL}
\begin{align}
f_0(q^2)&=\frac{{\cal L}_B}{1-q^2/M_0^2}
 \sum_{n=0}^{N-1}a_n^0 z(q^2)^n,\nonumber\\
f_i(q^2)&=\frac{{\cal L}_B}{1-q^2/M_*^2}
 \sum_{n=0}^{N-1}a_n^i
 \left[z(q^2)^n-(-1)^{n-N}\frac{n}{N}z(q^2)^N\right],
 \qquad i=+,T.
\label{eq:BCL-form-factors}
\end{align}
For $N=3$, $t_0=0$, and
$t_+=(m_B+m_K)^2$, the conformal variable is
\begin{equation}
 z(q^2)=\frac{\sqrt{t_+-q^2}-\sqrt{t_+-t_0}}
 {\sqrt{t_+-q^2}+\sqrt{t_+-t_0}}.
\label{eq:BCL-z}
\end{equation}
The coefficients used in the calculation are listed in
Table~\ref{tab:BCL-central-coefficients}, while the remaining BCL parameters
are given in Eq.~\eqref{eq:BCL-additional-inputs}.

\begin{table}[t]
\centering
\renewcommand{\arraystretch}{1.2}
\begin{tabular}{c c c c}
\hline\hline
$n$ & $a_n^0$ & $a_n^+$ & $a_n^T$ \\ \hline
0 &  0.254616273 &  0.254616273 &  0.254907841 \\
1 &  0.211016714 & -0.708471001 & -0.654938491 \\
2 &  0.026907767 &  0.309690152 &  0.362659041 \\ \hline\hline
\end{tabular}
\caption{Central coefficients of the $N=3$ BCL expansion used for the
$B\to K$ form factors.}
\label{tab:BCL-central-coefficients}
\end{table}

\begin{equation}
{\cal L}_B=1.303655672,\qquad
M_0=5.729495108\ \mathrm{GeV},\qquad
M_*=5.415766152\ \mathrm{GeV}.
\label{eq:BCL-additional-inputs}
\end{equation}
The tensor coefficients correspond to $f_T(q^2,\mu_b)$ at
$\mu_b=4.8\,\mathrm{GeV}$. The phenomenological use of these form factors in
$B\to K\ell^+\ell^-$ and $B\to K\nu\bar\nu$ is discussed in
Ref.~\cite{Parrott:2022sm}.
\section{Rare top decays}\label{sec:top-decays}
\begin{figure}[t!]
\centering
\begin{subfigure}{0.32\linewidth}
\centering
\raisebox{2.2ex}[\height][\depth]{%
\resizebox{\linewidth}{!}{%
\lqtreediagram{t}{e_\alpha^+}{c}{e_\beta^-}
}}
\end{subfigure}
\hfill
\begin{subfigure}{0.32\linewidth}
\centering
\resizebox{\linewidth}{!}{%
\begin{tikzpicture}[every node/.style={font=\small}]
  \begin{feynman}
    \vertex (t) at (-2.5,0) {$t$};
    \vertex[dot, minimum size=3pt] (vL) at (-1.2,0) {};
    \vertex[dot, minimum size=3pt] (vR) at (1.2,0) {};
    \vertex (c) at (2.5,0) {$c$};
    \vertex[dot, minimum size=3pt] (vS) at (0,1.3) {};
    \vertex (X) at (2.5,2.1) {$X=\gamma,g,Z$};
    \diagram*{
      (t) -- [fermion] (vL),
      (vR) -- [fermion] (c),
      (vL) -- [anti fermion, edge label'={$e_\alpha^C$}] (vR),
      (vL) -- [scalar, quarter left, edge label={$S_a$}] (vS),
      (vS) -- [scalar, quarter left] (vR),
      (vS) -- [boson] (X)
    };
  \end{feynman}
\end{tikzpicture}
}
\end{subfigure}
\hfill
\begin{subfigure}{0.32\linewidth}
\centering
\resizebox{\linewidth}{!}{%
\begin{tikzpicture}[every node/.style={font=\small}]
  \begin{feynman}
    \vertex (t) at (-2.5,0) {$t$};
    \vertex[dot, minimum size=3pt] (vL) at (-1.2,0) {};
    \vertex[dot, minimum size=3pt] (vR) at (1.2,0) {};
    \vertex (c) at (2.5,0) {$c$};
    \vertex[dot, minimum size=3pt] (vF) at (0,1.3) {};
    \vertex (X) at (2.5,2.1) {$X=\gamma,Z$};
    \diagram*{
      (t) -- [fermion] (vL),
      (vR) -- [fermion] (c),
      (vL) -- [scalar, edge label'={$S_a$}] (vR),
      (vL) -- [anti fermion, quarter left, edge label={$e_\alpha^C$}] (vF),
      (vF) -- [anti fermion, quarter left] (vR),
      (vF) -- [boson] (X)
    };
  \end{feynman}
\end{tikzpicture}
}
\end{subfigure}
\caption{Leptoquark contributions to rare top decays. The left panel shows the
tree-level exchange inducing $t\to c e_\alpha^+e_\beta^-$. The center and
right panels show representative one-loop diagrams for $t\to cX$, with the
gauge boson coupled to the leptoquark and the charged lepton, respectively.
Here $S_a$ denotes either physical leptoquark ($a=1,2$); $X=\gamma,g,Z$ in the
center panel, while $X=\gamma,Z$ in the right panel.}
\label{fig:top-decays}
\end{figure}
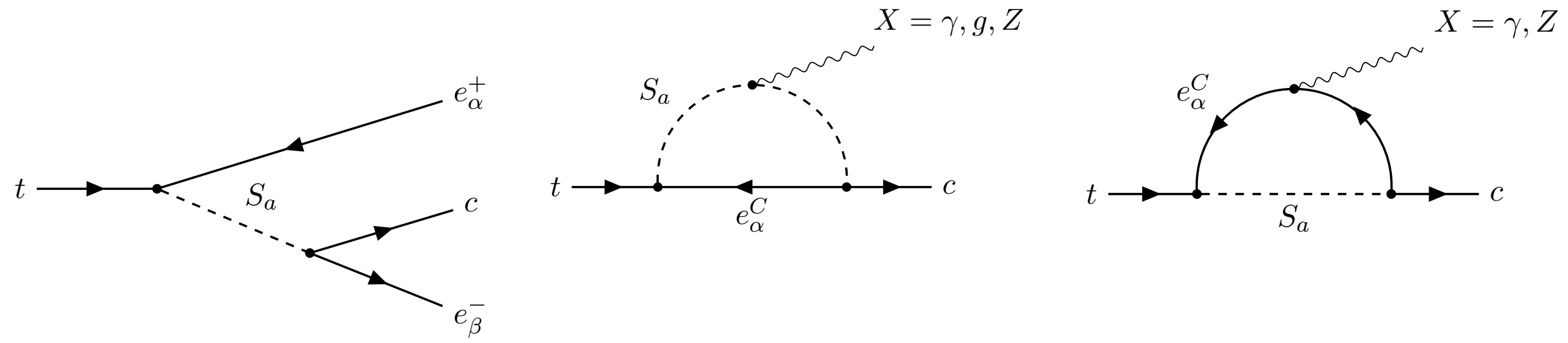

The charged-lepton interactions of the
model can induce rare top decays. Here we estimate their rates and show why they are negligible for the
$\widetilde Y^R=0$ benchmark adopted in the numerical analysis.
The tree-level topology for the three-body decay and the representative
one-loop topologies for the two-body channels are shown in
Fig.~\ref{fig:top-decays}.

The charged-lepton interaction in Eq.~\eqref{eq:alt-physical-yukawa} induces
$t\to c e_\alpha^+e_\beta^-$ at tree level. For the numerical estimates
below, we set $\widetilde Y^R=0$, consistently with the main analysis. For
$m_{S_1},m_{S_2}\gg m_t$ and neglecting the final-state fermion masses, its
branching fraction is approximately
\begin{equation}
 \mathcal{B}_{\alpha\beta}\equiv\mathcal{B}(t\to c e_\alpha^+e_\beta^-)
 \simeq\frac{m_t^5}{6144\pi^3\Gamma_t}
 |Y^L_{3\alpha}Y^{L*}_{2\beta}|^2
 \left(\frac{\cos^2\theta}{m_{S_1}^2}
 +\frac{\sin^2\theta}{m_{S_2}^2}\right)^2,
 \label{eq:top-branching-YL}
\end{equation}
where $m_t\simeq173\,\mathrm{GeV}$ and $\Gamma_t\simeq1.4\,\mathrm{GeV}$
~\cite{ParticleDataGroup:2026mpi}. The suppression is particularly strong when
$Y^L$ is reconstructed from neutrino masses using Eq.~\eqref{eq:alt-reconstruction}.
For example, $m_{S_1}=1\,\mathrm{TeV}$, $m_{S_2}=1.5\,\mathrm{TeV}$,
$\sin\theta=1/\sqrt{2}$ and $Y^R=\operatorname{diag}(0,1,1)$ give
$\max_\alpha|Y^L_{2\alpha}|=7.6\times10^{-9}$ and
$\max_\alpha|Y^L_{3\alpha}|=1.4\times10^{-9}$ for normal ordering with $m_1=0$.
The largest individual three-body branching fraction is then
$3.3\times10^{-41}$, far below experimental sensitivities of order
$10^{-7}$--$10^{-6}$~\cite{ATLAS:2024mutauTop}.

The two-body channels $t\to cX$, with $X=\gamma,g,Z$, arise at one loop
and are further suppressed, as illustrated in
the center and right panels of Fig.~\ref{fig:top-decays}. For
$\widetilde Y^R=0$, their leading
dipole amplitudes can be written as
\begin{equation}
\begin{aligned}
\mathcal M_X={}&\bar u_c i\sigma^{\mu\nu}q_\nu
 \left(A_L^X P_L+A_R^X P_R\right)u_t\,\epsilon_\mu^*,\\
A_R^X\simeq{}&\frac{g_X m_t}{16\pi^2}
 \sum_\alpha Y^L_{3\alpha}Y^{L*}_{2\alpha}
 \Bigg[\frac{c_\theta^2}{m_{S_1}^2}
 f_X\!\left(\frac{m_{e_\alpha}^2}{m_{S_1}^2}\right)+\frac{s_\theta^2}{m_{S_2}^2}
 f_X\!\left(\frac{m_{e_\alpha}^2}{m_{S_2}^2}\right)\Bigg],\\
A_L^X={}&\mathcal O\!\left(\frac{m_c}{m_t}\right)A_R^X,
\end{aligned}
\label{eq:top-two-body-amplitude}
\end{equation}
where $g_\gamma=e$, $g_g=g_s$, $g_Z=g/c_W$, and the dimensionless function
$f_X$ includes the diagrams with $X$ attached to the scalar and, for
$X=\gamma,Z$, to the charged lepton. They are of order unity in the
heavy-leptoquark limit. The corresponding dipole contribution is
\begin{equation}
\Gamma(t\to cX)=\frac{\kappa_X m_t^3}{16\pi}
 \left(1-\frac{m_X^2}{m_t^2}\right)^3
 \left(|A_L^X|^2+|A_R^X|^2\right),\qquad
\kappa_\gamma=\kappa_Z=1,\quad\kappa_g=C_F=\frac43.
\label{eq:top-two-body-width}
\end{equation}
With the reconstruction in Sec.~\ref{sec:framework},
$\zeta\zeta^\dagger$ is diagonal for the restricted texture, but
\begin{equation}
Y^L Y^{L\dagger}
=V^{*}(\zeta\zeta^\dagger)V^{T},
\label{eq:top-flavor-factor}
\end{equation}
is not diagonal in general. Therefore, the mass-independent flavor factor
$\sum_\alpha Y^L_{3\alpha}Y^{L*}_{2\alpha}
=(Y^L Y^{L\dagger})_{32}$ does not vanish exactly.
For the benchmark above, its modulus is only $5.8\times10^{-18}$, leading to
two-body branching fractions below approximately $10^{-41}$ for order-one
loop functions. Both the tree-level and one-loop rates are therefore
negligible, and rare top decays are not included as constraints in the
numerical analysis with $\widetilde Y^R=0$.

If the couplings in $\widetilde Y^R$ are allowed to be
of order unity, rare top decays can become moderately relevant for a light
leptoquark spectrum. For example, with the diagonal ansatz
$\widetilde Y^R=\operatorname{diag}(\widetilde y_u,\widetilde y_c,
\widetilde y_t)$, maximal mixing, and
$(m_{S_1},m_{S_2})=(1,1.5)\,\mathrm{TeV}$, the dominant channel gives
\begin{equation}
\mathcal{B}(t\to c\tau^+\mu^-)
\simeq2.97\times10^{-7}
|\widetilde y_t\widetilde y_c|^2,
\label{eq:top-diagonal-right-handed}
\end{equation}
which is below, but within the same order of magnitude as, the ATLAS bound
$\mathcal{B}(t\to\mu\tau q)<8.7\times10^{-7}$ at $95\%$ C.L.
~\cite{ATLAS:2024mutauTop}. Since the rate scales approximately as
$m_S^{-4}$, the same prediction falls to
$2.97\times10^{-11}|\widetilde y_t\widetilde y_c|^2$ for
$(m_{S_1},m_{S_2})=(10,15)\,\mathrm{TeV}$.

The loop factor additionally suppresses the two-body channels.
Moreover, for the diagonal ansatz the pure right-handed flavor factor
vanishes, while the remaining mixed terms involve the tiny reconstructed
$Y^L$ couplings and are therefore negligible. Thus, order-one
$\widetilde Y^R$ couplings can bring rare top decays close to current
sensitivity only for leptoquark masses near the TeV scale; the rates rapidly
become negligible for heavier spectra.

\nocite{apsrev41Control}
\bibliographystyle{apsrev4-1}
\bibliography{refs}

\end{document}